\documentclass{amsart}
\pdfoutput=1
\usepackage{amsaddr}
\usepackage{amssymb}
\usepackage{multirow}
\usepackage{booktabs}
\usepackage{graphicx}
\usepackage[shortlabels]{enumitem}
\usepackage{xr}
\usepackage{url}
\usepackage{subfig}
\usepackage[nobysame]{amsrefs}

\theoremstyle{plain}%
\newtheorem{theorem}{Theorem}
\newtheorem{lemma}{Lemma}
\newtheorem{corollary}{Corollary}

\theoremstyle{remark}%
\newtheorem{remark}{Remark}%

\theoremstyle{definition}%
\newtheorem{definition}{Definition}
\newtheorem{assumption}{Assumption}
\begin{document}

\title[A Robust Framework for Graph-based Two-Sample Tests]{A Robust Framework for Graph-based Two-Sample Tests Using Weights}

\author{Yichuan Bai and Lynna Chu}

\address{Department of Statistics, Iowa State University, Ames, Iowa, USA}

\begin{abstract}
Graph-based tests are a class of non-parametric two-sample tests useful for analyzing high-dimensional data. The test statistics are constructed from similarity graphs (such as $K$-minimum spanning tree), and consequently, their performance is sensitive to the structure of the graph. When the graph has problematic structures (for example, hubs), as is common for high dimensional data, this can result in low power and unstable performance among existing graph-based tests. We address this challenge by proposing new test statistics that are robust to problematic structures of the graph and can provide reliable inferences. We employ an edge-weighting strategy using intrinsic characteristics of the graph that are computationally simple and efficient to obtain. The limiting null distribution of the robust test statistics is derived and shown to work well for finite sample sizes. Simulation studies and data analysis of Chicago taxi-trip travel patterns demonstrate the new tests' improved performance across a range of settings.
\end{abstract}


\maketitle

\section{Introduction}
We focus on testing the equality of distributions for observations in the high dimensional setting, where the dimension of the observation $d$ may be much larger than the sample size $N$. 
Suppose we have two samples $$\{\mathbf{X_1, \hdots, X_{n_1}}\}\text{ and }\{\mathbf{Y_1, \hdots, Y_{n_2}}\}$$ of $d$-dimensional observations that are independently and identically distributed from unknown distributions $F_X$ and $F_Y$, respectively. The two-sample problem aims to test  $H_0: F_X = F_Y$ against an omnibus alternative $H_1: F_X \neq F_Y$. This is a classic statistical problem but made more challenging by the increasing complexity of modern data, where observations can be high-dimensional data objects ($d >> N$). In this setting, it is often intractable to express or estimate $F_X$ and $F_Y$ directly due to the curse of dimensionality. Substantial developments have been made by the contemporary statistics community to address such challenges. For example, non-parametric two-sample tests for multivariate and high-dimensional data have been proposed using distances (\cite{baringhaus2004new, szekely2004testing, biswas2014nonparametric, li2018asymptotic}), generalized ranking (\cite{liu1993quality, hall2002permutation}), and kernels (\cite{gretton2012kernel, song2020generalized, Zhu2021}).

While all of the mentioned methods can be applied to the high-dimensional setting, many do not explicitly address how to resolve various aspects of the curse of dimensionality. For example, distance-based test statistics are commonly used in the high-dimensional setting, but it has been observed that distances may not be meaningful in high-dimensional space since they have a tendency to concentrate when $d$ is large. As such, distance-based test statistics may have trouble effectively distinguishing similarities between observations, leading to reduced power. Moreover, the distribution of distances becomes considerably skewed as dimensionality increases, resulting in a phenomenon known as \textit{hubness}. To be precise, let $N_k(x)$ be the number of times an observation $x$ is among the $k$ nearest neighbors of all other points in the data set. When the dimensionality is high, the distribution of $N_k$ becomes right-skewed, resulting in the emergence of hubs. This hubness phenomenon affects methods that directly (or indirectly) make use of distances between observations; this includes pairwise distance-based tests such as the energy statistic (\cite{szekely2004testing}) and graph-based tests based on interpoint distances (described below). As a result, many existing two-sample tests are often vulnerable to the hubness aspect of the dimensionality curse, which can incur poor or unstable performance under various scenarios. 

In this paper, we explore the hubness phenomenon and its effect on a class of tests based on geometric graphs constructed using interpoint distances. We refer to these as graph-based two-sample tests; the first test was proposed by \cite{ori1979}, and since then, numerous extensions and theoretical developments have been made. For example, \cite{schilling1986multivariate} and \cite{henze1988multivariate}, \cite{rosenbaum2005exact}, and \cite{biswas2014distribution}  proposed test statistics specifically for $k$-NN graphs, minimum distance pairing, and Hamiltonian graphs, respectively. \cite{gen2017}, \cite{wei2018}, and \cite{max2019} proposed new graph-based test statistics that target a wider range of alternatives. \cite{banerjee2020nearest} and \cite{banerjee2023bootstrapped} proposed modifications of graph-based tests targeting the setting when heterogeneity is present in the two samples due to latent subpopulations. \cite{zhou2023rank} proposed incorporating ranks in a similarity graph to boost the power of existing tests. \cite{zhu2023limiting} studied asymptotic results for dense graphs. Details on constructing graph-based tests are provided in Section \ref{sec:overview}.

Despite their utility, these graph-based tests are sensitive to problematic data structures that can arise in the graphs. If the similarity graph is relatively flat, the existing tests work quite well. However in the presence of hubs and other problematic structures, the current tests suffer from reduced power and unreliable inference. We illustrate and explain why the hubness phenomenon can cause complications for the existing graph-based tests in Section \ref{sec:limitation}. While some graph-based methods, such as the cross-match test based on non-bipartite matching \cite{rosenbaum2005exact} and the Shortest Hamiltonian Path (SHP)-based test  \cite{biswas2014distribution}, can mitigate the hubness problem by placing constraints on the graph construction, these tests tend to suffer from low power under some common scenarios when the observations are high-dimensional (see Appendix \ref{Appendix:con_graph} for additional details). Recently, \cite{zhu2023robust} proposed a graph generation method that also places constraints on the graph; their approach involves optimizing an objective function with a penalty for a large node degree. However, if the similarity graph is constructed from domain knowledge or directly observed, as is often the case in real applications, their approach is no longer directly applicable. Moreover, their graph generating process could be computationally expensive and may also destroy vital internal connections between observations in the similarity graph. If the hubness is too extreme, their approach deletes the hub from the graph. As demonstrated in Section \ref{sec:application}, identifying the problematic hub may be nuanced, and straightforward deletion may not be ideal. 

To address the hubness problem while preserving power and similarity information, we take a different approach and propose a robust framework for graph-based tests that employs an edge weighting strategy on the graph-based test statistics. We do not place constraints on graph construction nor generate a new graph, but instead use weights that are derived from the intrinsic characteristics of the similarity graph. These weights can mitigate the influence of hubs while effectively retaining power in the presence of hubs. We demonstrate through theoretical analysis, simulation studies, and real data applications the improved performance of this robust framework.

The paper is organized as follows. In Section \ref{sec:framework}, we review the graph-based testing framework and discuss the hubness phenomenon. We then propose a robust solution in Section \ref{sec:method}, which involves choosing weights to dampen the effect of hubs. In Section \ref{sec:asymptotic}, the asymptotic null distributions of the proposed test statistics are derived. Section \ref{sec:performance} examines the power of the robust test statistics under different simulation settings. In Section \ref{sec:application}, the robust test statistics are illustrated in the analysis of Chicago taxi data and some concluding remarks are given in Section \ref{sec:conclusion}.

\section{Graph-based Testing Framework}\label{sec:framework}

\subsection{Background}\label{sec:overview}

Graph-based tests provide a general framework to conduct two-sample tests for multivariate and non-Euclidean data. A similarity graph is constructed from the pooled observations of both samples according to a similarity measure (such as Euclidean distance). The similarity graph can be constructed based on a certain criterion. For example, a minimum spanning tree (MST) is a similarity graph that connects all observations in such a way that the total distance across edges is minimized. A $k$-MST is the union of MST and $k-1$ spanning trees, where the $i$th $(i > 1)$ spanning tree does not contain any edges from the first $i-1$ spanning trees. Other examples include the $k$-nearest neighbor graph ($k$-NNG), where each observation is connected to its $k$ nearest neighbors. Alternatively, the graph could be constructed according to domain knowledge and expertise. 

Three quantities of the graph are computed: the number of edges connecting between the two samples ($R_0$), the number of edges connecting within sample $\textbf{X}$ ($R_1$), and the number of edges connecting within sample $\textbf{Y}$ ($R_2$). A combination of these edge counts is used to construct different graph-based test statistics. \cite{ori1979} proposed using (a standardized) $R_0$ as the test statistic such that a small $R_0$ is evidence against the null hypothesis that the two distributions are equal. Their rationale was that if the two samples really do come from different distributions, then the number of edges connecting between different samples should be relatively small. While a small $R_0$ as evidence against the null holds well when the two distributions differ in means, this rationale can be invalid for more general alternatives - for example, when the change in distribution also involves scale change or the two samples are unbalanced. To resolve this, graph-based test statistics were proposed in \cite{gen2017}, \cite{wei2018}, and \cite{max2019} that use a combination of $R_1$ and $R_2$ and can target a wider range of alternatives. 

\subsection{Hubness phenomenon in high-dimensional data}
\label{sec:motivation}

\begin{figure}[!t]
\centering
\includegraphics[width=\linewidth]{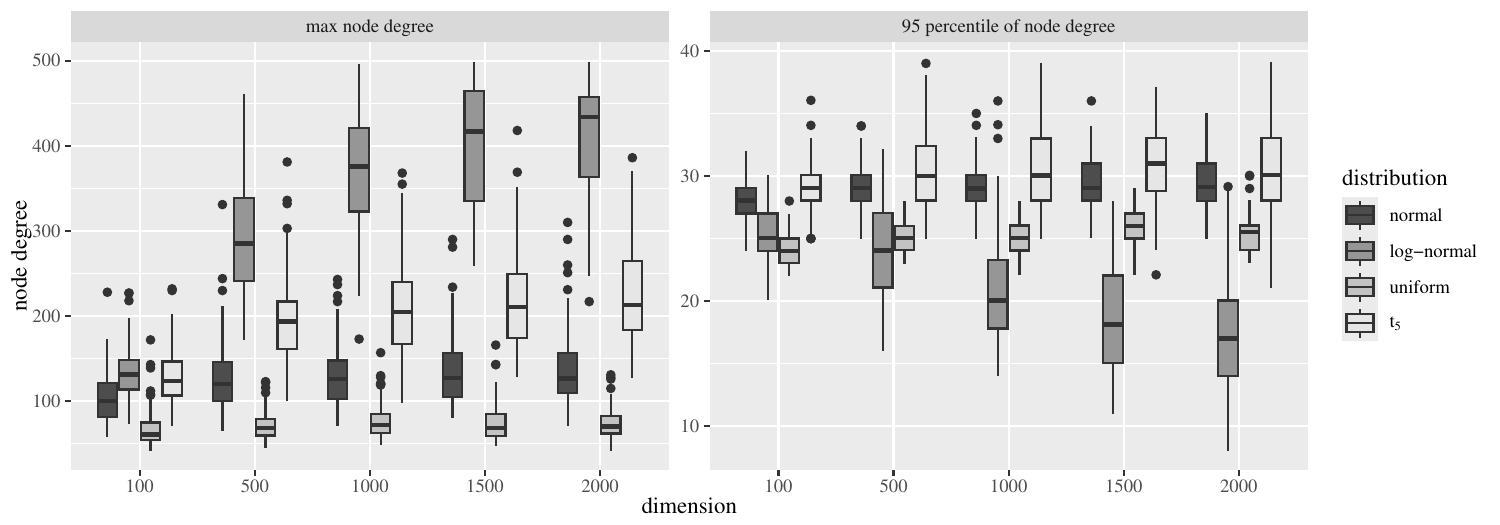}
\caption{Boxplot of maximum and 95th percentiles of node degrees for different dimensions. Results are from 100 simulations with $n = 500$, where observations are drawn from $d$-dimensional normal, log-normal, uniform, and t distributions.}
\label{fig:max_nodedegree}
\end{figure}

Hubs, defined to be nodes in the graph with a large degree, are a product of the curse of dimensionality. The hubness phenomenon was carefully studied in \cite{radovanovic2010hubs}, which showed that hubs are an inherent property of data distributions in high-dimensional and not an artifact of finite samples or specific data distributions. Their theoretical analysis showed that the probability a hub emerges increases as the data dimension increases. The high-dimensional setting amplifies the tendency of central observations (observations close to the mean) to become hubs, effectively making it easier for such an observation to become a `popular' or `central' node. As a result, $k$-MSTs and $k$-NNGs constructed on high-dimensional data tend to have large hubs under standard distance measures, such as $L_p$. To see that the presence of hubs is a common phenomenon for high-dimensional data, we construct $5$-MST graphs using Euclidean distance and report the maximum and 95th percentile of node degrees. As shown in Figure \ref{fig:max_nodedegree}, we see that the maximum node degrees are more than three times as much as the 95th percentiles. Similar results using $5$-NN constructed from Euclidean distance are shown in Appendix \ref{Appendix:max_nodedegree_nn}. Clearly it is not uncommon to have a node with a degree much larger than the majority of other nodes' in the similarity graph. These hubs can be highly influential nodes and can distort final inference results depending on whether these observations are included or excluded in graph construction.

\subsection{Limitations of current graph-based tests}\label{sec:limitation}
Consider the following example that illustrates why large hubs may cause problems in the existing graph-based tests. In Scenario 1, we generate two samples $(n_1 = n_2 = 200$) with moderate dimension that differ in mean and variance: $$\mathcal{F}_1:\mathcal{N}(\textbf{0}_{s}, \textbf{I}_{s}) \text{ and } \mathcal{F}_2: \mathcal{N}(\boldsymbol{\sqrt{(0.2log(s)/s)}1_{s}}, (1+3log(s)/s)^2\textbf{I}_{s} );$$
in Scenario 2, the samples ($n_1 = n_2 = 200$) are generated with the same change in mean and variance but the observations are high-dimensional: $$\mathcal{F}_1:\mathcal{N}(\textbf{0}_{d}, \textbf{I}_{d}) \text{ and } \mathcal{F}_2: \mathcal{N}(\boldsymbol{(\sqrt{(0.2log(s)/s)}1_{s}, 0_{d-s})}, \big(\begin{smallmatrix}
  (1+3log(s)/s)^2\textbf{I}_{s} & \textbf{0}\\
  \textbf{0} & \textbf{I}_{d-s}
\end{smallmatrix}\big)),$$
 where $d = 1000$ and $s = \lfloor{\sqrt{d}}\rfloor$.  A 5-MST is constructed from Euclidean distance on the pooled observations ($n_1 + n_2 = 400)$. We observe that the maximum node degree of the graph is 64 in Scenario 1 and 80 in Scenario 2. The generalized edge-count test $S$ (\cite{gen2017}) and the max-type edge-count test $M$ (\cite{max2019}), which consist of different combinations of $R_1$ and $R_2$, are applied to both scenarios since they can detect general distributional differences (i.e., both mean and variance change). A large value of $S$ or $M$ serves as evidence against the null hypothesis. Both tests are capable of detecting the difference between two samples under Scenario 1. However, in Scenario 2, in the presence of a larger hub, both tests cannot reject the null hypothesis at the 10\% significance level, with p-values 0.3899 and 0.2815, respectively.

\begin{table*}[ht]
	\caption{Graph-based quantities for 5-MST under Scenario 1 and 2.}
	\centering
	\begin{tabular*}{\columnwidth}{@{\extracolsep{\fill}}lcccccc@{\extracolsep{\fill}}}
		\toprule
		& $R_1$ & E($R_1$) & Var($R_1$) & $R_2$ & E($R_2$) & Var($R_2$) \\
		\midrule
    	Scenario 1 &	 956&497.5 &1978.756&144 &497.5 &1978.756 \\	
    	Scenario 2 &571 & 497.5  & 2872.114 & 431&497.5 & 2872.114   \\
		\bottomrule
	\end{tabular*}
	\offinterlineskip
	\label{tab:table 2}
\end{table*}

\begin{figure}[!t]
\centering
\includegraphics[width=0.8\linewidth]{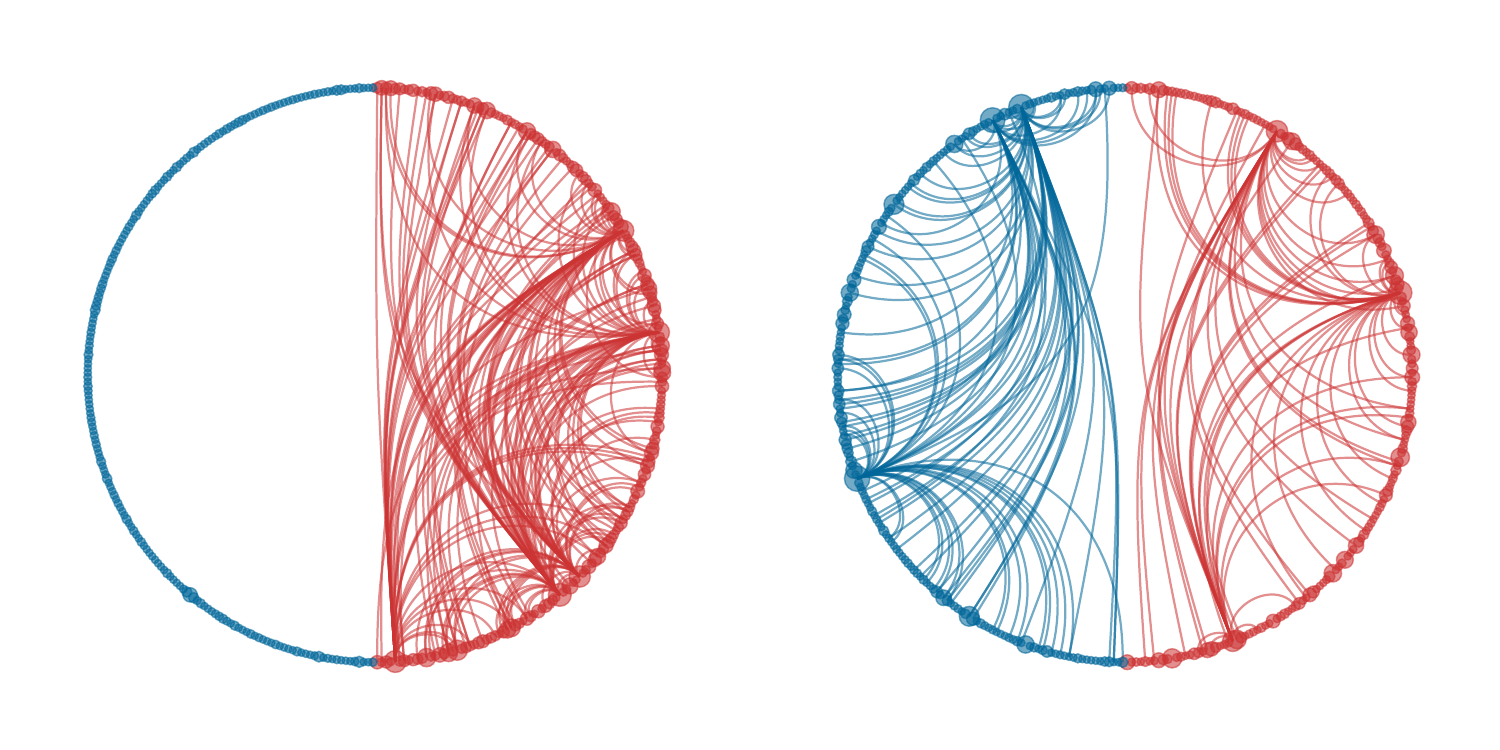}

\caption{Illustration of edges connected to hubs (defined in this setting to be nodes with node degree larger than 50) in the similarity graph for Scenario 1 (left) and Scenario 2 (right). Hubs from Samples 1 and 2 are represented by red and blue points, respectively, along the circle perimeter. The size of the point corresponds to the node degree. Edges connecting observations from Sample 1 are in red and from Sample 2 are in blue.}
\label{fig:illustrate}
\end{figure}

Table~\ref{tab:table 2} sheds insight into why this happens. Two graph-based within-sample edge counts $R_1$ and $R_2$ and their expectations under the null $E(R_1)$ and $E(R_2)$ are reported. 
Under the alternative, we would expect the absolute value of differences between the within-sample edge counts and their null expectations to be relatively large ($|R_1 - E(R_1)|$ and/or $|R_2 - E(R_2)|)$. Figure \ref{fig:illustrate} illustrates how the edge counts behave in the two scenarios. We plot only those edges that are connected to hubs - which we define in this setting to be any node with a degree larger than 50. 

We can see in Scenario 1, $R_1$ and $R_2$ behave as we expect. Observe that in Figure \ref{fig:illustrate}, hubs are generated in Sample 1 with many within-sample connections (so we can see many red edges), making $R_1$ large. On the other hand, most of the observations in Sample 2 (with a larger variance) connect to those in Sample 1, making $R_2$ small (we do not see any blue edges). Then, the differences between $R_1$ and $R_2$ and their respective expectations are relatively large as shown in the first row of  Table~\ref{tab:table 2}, and it follows that existing tests have the power to reject the null. 

In Scenario 2, two problems arise in the presence of a hub. First, hubs tend to form in both samples. A hub in the sample with a larger variance will form edges with observations from the same sample, which increases $R_2$ (we see more blue edges in Figure \ref{fig:illustrate}), and form between-sample edges (decreasing $R_1$). The relative differences between the edge counts and their respective expectations in Scenario 2 become smaller as shown in the second row of  Table~\ref{tab:table 2}, causing both tests to lose power. Second, the variances of both $R_1$ and $R_2$ increase, further inhibiting the power of the test statistics. This second problem can be clearly seen by studying the analytical expression of the variance of the edge counts, $R_j$ $j = 1, 2$, under the permutation null distribution. Let $G$ denote the similarity graph and its set of edges, $|G|$ denote the number of edges in $G$, $G_i$ be the subgraph including all edge(s) that connect to node $i$, and $|G_i|$ be the degree of node $i$ in $G$. The variance expression of $R_j$ is:
\begin{equation*}
\begin{split}
\text{Var}(R_j) = & [2C(N-n_j)+|G|(|G|-1)(n_j-3)]\frac{n_j(n_j-1)(n_j-2)}{N(N-1)(N-2)(N-3)} + \\
&\mu_j(1- \mu_j) ,
\end{split}
\end{equation*}
where $j = 1, 2$, $\mu_j = E(R_j) = |G|\frac{n_j(n_j-1)}{N(N-1)}$, $C = \frac{1}{2}\sum_{i = 1}^N|G_i|^2 -|G|$, $n_j$ is the number of observations in sample $j$, and $N = n_1+ n_2$. In the presence of a hub, both $\sum_{i = 1}^N |G_i|^2$ and $C$ increase, where $C$ represents the number of edge pairs sharing a common node, which in turn results in an inflated variance for $R_1$ and $R_2$.

\subsection{Our Contribution} 
When the size or density of hubs is large, existing graph-based tests can suffer from limited power and unstable performance. We propose new test statistics that are useful even in the presence of hubs. Specifically, we propose to apply appropriate weights to the test statistics that will dampen the effect of hubs while still retaining crucial similarity information. We show that these weights can improve power and resolve the variance boosting problem in the presence of problematic graph structures. We provide recommendations for weights as a function of node degrees and demonstrate that these work well in a range of scenarios. The limiting null distribution of these new robust test statistics is derived under mild conditions on the weights, and we show that the limiting distribution is quite accurate for finite sample sizes. Unless stated otherwise, we use the 5-MST constructed from $L_2$ distances of the pooled observations as the similarity graph in simulations. The robust edge-count tests can be implemented using the R package `rgTest'. Code for simulations and our application is available  at \url{https://github.com/stat-yb/robustEtest.git}.

\section{Robust edge-count test statistics}\label{sec:method} 

Our approach is to flatten the similarity graph in order to limit the influence of hubs without incurring too much of a loss of similarity information so that the testing procedure can still retain power. To do so, we propose to apply weights that are functions of the graph's node degrees to the edge-count test statistics. The weights should be designed such that edges connected to a hub are down-weighted, while other edges are left mostly undisturbed. Let $d_i$ denote the node degree of node $i$ in a graph $G$. Let $(i, j)$ represent the edge connecting observations $i$ and $j$ in graph $G$. Let $w_{ij}$ denote the weight on edge $(i, j)$ where $w_{ij}$ is the value of the weight function $W(d_i,d_j)$, with $W(d_i,d_j)$ defined to be a function of $d_i$ and $d_j$. Discussions about the choice of weight functions are deferred to Section \ref{sssec:weights}. 

We apply weights $w_{ij}$ to the edge counts $R_1(i,j)$, and $R_2(i,j)$, such that each edge  $(i,j) \in G$ is weighted by a combination of $d_i$ and $d_j$. Let $g_i = 0$ if the observation $i$ is from Sample $\boldsymbol{X}$, and 1 otherwise. Let $n_1$ be the sample size of Sample $\boldsymbol{X}$, $n_2$ be the sample size of Sample $\boldsymbol{Y}$, and $N = n_1 + n_2$. We define
\begin{equation*}
R_1^w = \sum_{(i, j) \in G}R_1(i,j), \ R_2^w = \sum_{(i, j) \in G}R_2(i,j).
\end{equation*}
where $R_1(i,j) = w_{ij} I(J_{(i, j)}=1),\ R_2(i,j) =  w_{ij} I(J_{(i, j)}=2)$, and
\begin{equation*}
J_{(i, j)} = \begin{cases}
1 &\text{if $g_i = g_j = 0$},\\
2 &\text{if $g_i = g_j = 1$}. 
\end{cases}
\end{equation*}
The robust generalized edge-count test statistic is defined to be:
\begin{equation*}
 S_R = (R_1^w - \mu_1^w,  R_2^w - \mu_2^w)(\Sigma^w)^{-1}\begin{pmatrix}
  R_1^w - \mu_1^w\\ 
  R_2^w - \mu_2^w
\end{pmatrix},
 \end{equation*}
where $ \mu_1^w= E(R_1^w), \mu_2^w= E(R_2^w)$, and
\begin{equation*}
\Sigma^w = \begin{pmatrix}
  \text{Var} (R_1^w) & \text{Cov} (R_1^w, R_2^w)\\ 
  \text{Cov} (R_1^w, R_2^w) & \text{Var} (R_2^w)
\end{pmatrix} = \begin{pmatrix}
  \Sigma_{11} & \Sigma_{12}\\ 
  \Sigma_{21} & \Sigma_{22}
\end{pmatrix}.
\end{equation*}

\begin{theorem} \label{th:decompose}
$S_R$ can be expressed as 
$$S_R = (Z^R_\text{diff})^2 + (Z^R_{w})^2,$$
with $\text{Cov}(Z^R_\text{diff},Z^R_{w}) = 0$,
where
$$Z_\text{diff}^R = [(R_1^w - R_2^w) - E(R_1^w - R_2^w)]/[\text{Var}(R_1^w - R_2^w)]^{1/2},$$ $$Z_w^R= [(qR_1^w + pR_2^w) - E(qR_1^w + pR_2^w)]/[\text{Var}(qR_1^w + pR_2^w)]^{1/2},$$
$p = (n_1-1)/(N-2)$, and $q = 1- p$.
\end{theorem}
The proof of Theorem \ref{th:decompose} can be found in the Appendix \ref{Appendix:decompose}. 

Theorem \ref{th:decompose} leads us to propose the robust max-type edge-count test statistic: $$M_R = \text{max}(Z_w^R, |Z_\text{diff}^R|).$$

If the graph is relatively flat and no hub is present, then $d_i$ is similar for all $i \in [1, N]$, and the weights have little effect. However, in the presence of a problematic hub(s), the weights control the influence of edges connected to the hub, resulting in improved and reliable performance. This creates a test statistic that is increasingly robust to the underlying similarity graph and also resolves the variance boosting problem. 

The analytic expressions of expectations and variances involved above can be obtained by combinatorial analysis under the permutation null distribution. We present them in the following lemma. 

\begin{lemma} \label{th:meanvarori}
Under the permutation distribution, we have:
\begin{equation*}
\begin{split}
\mu_1^w =& \sum_{(i, j) \in G}w_{ij} \frac{n_1(n_1-1)}{N(N-1)},\ 
\mu_2^w = \sum_{(i, j) \in G}w_{ij} \frac{n_2(n_2-1)}{N(N-1)},\\
\Sigma_{11} =&[-S_2 +\frac{2(2N-3)}{N(N-1)}S_3+\frac{N-3}{n_2-1}\left(S_1+S_2\right)-\frac{4(N-3)}{N(n_2-1)}S_3 ] D_N,\\
 \Sigma_{22} =&[-S_2+ \frac{2(2N-3)}{N(N-1)}S_3 +\frac{N-3}{n_1-1}\left(S_1+S_2\right)-\frac{4(N-3)}{N(n_1-1)}S_3] D_N,\\
 \Sigma_{12} = &[-S_2 +\frac{2(2N-3)}{N(N-1)}S_3]D_N,
\end{split}
\end{equation*}
where $S_1 = \sum_{(i, j) \in G}w_{ij}^2$, $S_2 = \sum_{(i, j), (i, k) \in G} w_{ij}w_{ik}$, $S_3 = \sum_{(i, j), (k, l) \in G}w_{ij}w_{kl}$ and $D_N = [n_1n_2(n_1-1)(n_2-1)]/[N(N-1)(N-2)(N-3)]$.
\end{lemma} 

The proof of this lemma can be found in the Appendix \ref{Appendix:mean_var}. Using the results from Lemma \ref{th:meanvarori}, the expectations and variances involved in $Z_\text{diff}^R$ and $Z_w^R$ can be obtained as follows:
\begin{equation*}
\begin{split}
 &E(R_1^w - R_2^w) = \sum_{(i, j) \in G}w_{ij} \frac{n_1 - n_2}{N},\ E(qR_1^w+pR_2^w) = \sum_{(i, j) \in G}w_{ij} \frac{(n_1-1)(n_2-1)}{(N-1)(N-2)},\\
  &\text{Var}(R_1^w - R_2^w)  = 
  [ (S_1+S_2)-\frac{4}{N}S_3] \frac{n_1n_2}{N(N-1)},\\
    &\text{Var}(qR_1^w+pR_2^w)  = [\frac{N-3}{N-2}S_1 -\frac{S_2}{N-2}  + \frac{2S_3}{(N-1)(N-2)}]
 D_N.
\end{split}
\end{equation*}

Large values of $S_R$ and $M_R$ are evidence against the null hypothesis of no distributional difference. The constructions of both $S_R$ and $M_R$ allow them to be powerful for general alternatives. When there is a change in the mean, both $R_1^w$ and $R_2^w$ tend to be larger than their null expectation - it follows that $Z^R_w$ will be large, which leads to a large $S_R$ and $M_R$. When a change in variance is present, without loss of generality, suppose the sample with the smaller variance is sample $\boldsymbol{X}$. Then $R_1^w$ is relatively large compared to its null expectation while $R_2^w$ is relatively small. In this case, $|Z^R_\text{diff}|$ tends to be large, which also leads to a large $S_R$ and $M_R$. The robust test statistics $S_R$ and $M_R$ default to the tests proposed in $S$ and $M$ when $w_{ij} = 1$ for all $(i,j)$. In $S$ and $M$, each edge has an equal contribution to the test statistic so that even those edges connected to problematic hubs are treated with the same weight as those that are not. By placing weights on the edges, we dampen the influence of hubs and effectively flatten the graph.

\begin{remark}\label{rm:welldefine}
The test statistics are well-defined under the following conditions: 
\begin{enumerate}
\item[(a.)] $\sum_{\{j, \text{s.t.} (i, j)\in G\}}w_{ij}$ are not all equal for all $i \in [1, N]$; 
\item[(b.)] $(N-3)S_1-S_2+\frac{2}{N-1}S_3 >0$,
where $S_1 = \sum_{(i, j) \in G}w_{ij}^2$, \\
$S_2 = \sum_{(i, j), (i, k) \in G} w_{ij}w_{ik}$ and $S_3 = \sum_{(i, j), (k, l) \in G}w_{ij}w_{kl}$. 
\end{enumerate}
\end{remark}
The proof of this remark can be found in the Appendix \ref{Appendix:welldefine}. For example, for a completely flat graph (all nodes have the same degree), then $w_{ij} = w_{i'j'},\ \forall i\neq j\neq  i'\neq j'\in [1, N]$ and $Z^R_\text{diff}$ is not well-defined. For a star-shaped graph, in which all observations connect to the same node, $Z^R_w$ is not well-defined.  Theorem \ref{th:bounds} ensure  $R_j^w, j = 1,2$ does not vanish to zero when the sample size goes to infinity. The proof of this theorem can be found in the Appendix \ref{Appendix:bounds}. 

\begin{theorem} \label{th:bounds}
Let $W$ be a weight function such that $W(i,j) = w_{ij}$, $\forall (i,j)\in G$. If the weight function $W$ is asymptotically bounded below by $1/|G|$ as $N\rightarrow \infty$, then $ \lim_{N\to\infty}R_s^w > 0$, for $s = 1, 2$.
\end{theorem}

\subsection{Proposed Weights} \label{sssec:weights}

The test statistics are defined for general weights that are functions of the node degrees and monotonically decreasing, as defined below.

\begin{definition}
A bivariate function is called monotonically decreasing if for all $x_1$,  $x_2$ and $y_0$ such that $x_1<x_2$, then $f(x_1, y_0) >f(x_2, y_0)$; and for all $y_1$,  $y_2$ and $x_0$ such that $y_1<y_2$, then $f(x_0, y_1) >f(x_0, y_2)$.
\end{definition}

In practice, users have the flexibility to choose their weights, provided that the test statistics are well-defined given the conditions in Remark \ref{rm:welldefine}. Since we allow the graph to be general, and the weights are properties of the graph, obtaining optimal weights for general similarity graphs is challenging, and we reserve this line of theoretical analysis for future work. Instead, we provide recommendations for data-driven weights based on empirical studies. We recommend a weight that (1) demonstrates reasonable power and (2) meets the conditions for our asymptotic theory. 

For edge $(i, j)$, we recommend the following weight function: 
\begin{equation} \label{eq:W}  W(d_i, d_j) = \frac{1}{\text{max}(d_i, d_j)}. \end{equation}
The weight function $W$ is bounded below by $1/|G|$ asymptotically and monotonically decreasing.

\begin{figure}[!t]
\centering
\includegraphics[width=0.65\columnwidth]{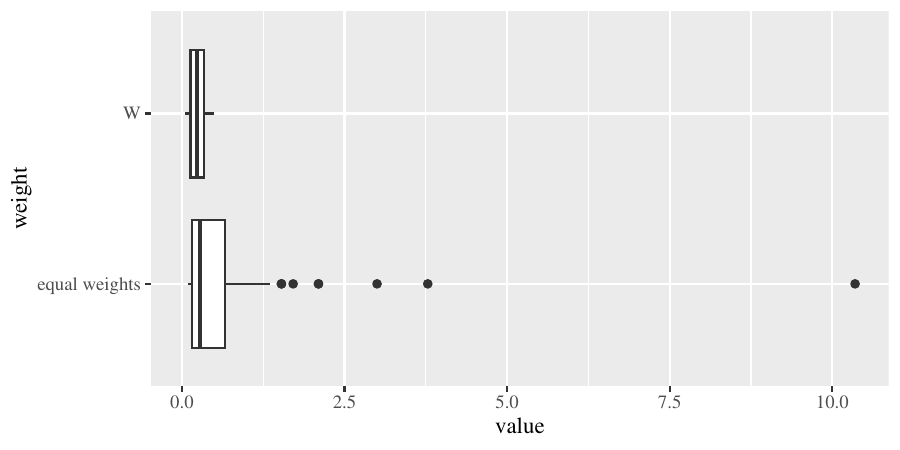}
\caption{The boxplots of $\sum_{ \{j, k: j \neq k, (i, j), (i, k) \in G\}} w_{ij}w_{ik}$ for $ i \in [1, N]$ using equal weights and weighted function $W$.} 
\label{fig:uni}
\end{figure}

We present the following examples to demonstrate how the weight function works in the robust test and its utility. First, we present an example to show how weights can temper the impact of hubness on the variance. A dataset with 100 observations is simulated from a 100-dimensional uniform distribution. According to Lemma \ref{th:meanvarori}, the change in the variance of the test statistics is contingent on the change in $S_1$, $S_2$, and $S_3$ for different similarity graph structures. When applying equal weights (which effectively treats all the edges as equal since $w_{ij}$ equals a constant $c$ for all $(i,j) \in G$), $S_1$ and $S_3$ are constant given a fixed number of edges, and any hubness in the similarity graph only affects $S_2$. In Figure \ref{fig:uni}, the boxplots of $\sum_{ \{j, k: j \neq k, (i, j), (i, k) \in G\}} w_{ij}w_{ik}$ for $ i \in [1, N]$, which is the dominant component in $S_2$, compares this quantity under equal weights and the weight function $W$ (\ref{eq:W}). There are several observations that form hubs in this setting. When using equal weights, it is clear that these hubs are still present and the variance boosting problem continues to manifest itself with large values of $\sum_{ \{j, k: j \neq k, (i, j), (i, k) \in G\}} w_{ij}w_{ik}$. On the other hand, when applying the weight function, the impact of the hubs is well-controlled.

To more comprehensively evaluate the performance of the weight function, we simulate 500 replications of two samples with $n_1 = 100$, $n_2 = 100$, and $d = 400$ from the $d$-dimensional log-normal distribution. The difference between the two samples is reflected by $\Delta_{\mu}$, where $\mu$ is the expected value of the variable's natural logarithm. The difference is equal across all dimensions such that $||\Delta_{\mu}||^2 = 2.5$. We record the maximum node degree of the similarity graph in each simulation, and group the simulations according to their maximum node degrees from low to high by each tenth percentile. The increase in the maximum node degree is indicative of a more severe hubness phenomenon.

\begin{figure}[ht!]
\centering
\includegraphics[width=\columnwidth]{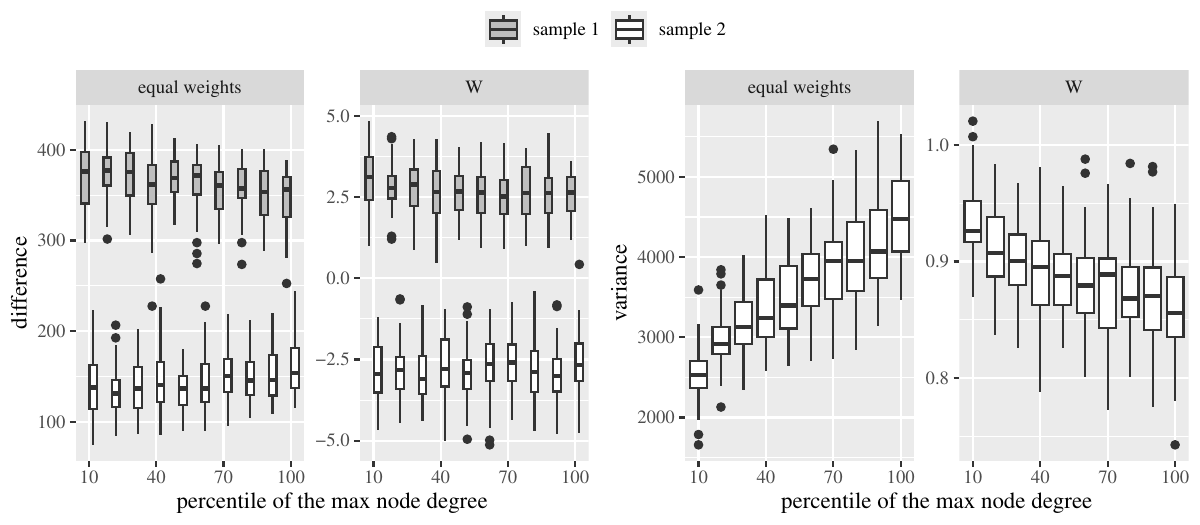}
\caption{Boxplots of $R_j^w - E(R_j^w)$ (left) and $\text{Var}(R_j^w)$ (right), $j = 1,2,$ using weights ($W$) compared to boxplots of corresponding quantities using equal weights. Simulations are grouped according to the percentile of the max node degrees. Only variances of sample 2 are presented since the sample sizes are equal, and the variances are roughly the same for both samples. }
\label{fig:lnorm}
\end{figure}

Figure \ref{fig:lnorm} presents boxplots of the difference between the robust within-sample edge counts and their expectations and the variances of robust within-sample edge counts. We compare these with their corresponding quantities using equal weights. Under the alternative, we anticipate a relatively large difference between the within-sample edge counts and their expectations. Under equal weights, as the maximum node degree increases, the relative difference decreases in Sample 1. However, the boxplots using $W$ do not exhibit a similar trend in Sample 1, which suggests higher power. Under equal weights, the variance 
also increases as the node degree of the hub increases. On the other hand, it is clear the weight function $W$ controls the variance from increasing.

\section{Asymptotic null distribution}\label{sec:asymptotic}
The robust edge-count test statistics are computationally straightforward to calculate and their significance can be obtained via resampling from the permutation distribution. However, as the sample size increases, permutation becomes increasingly computationally prohibitive. To make the tests practical for modern data sets, we study the limiting distributions of the robust edge-count test statistics. 

We define
\begin{itemize}
    \item $A_{(i, j)} = \{(i, j)\}\cup \{(i', j')\in G, (i, j)$ and $(i', j')$ share a node\},
    \item $B_{(i, j)} = A_{(i, j)} \cup \{(i'', j'')\in G, \exists (i', j') \in A_{(i, j)}$, such that $(i', j')$ and $(i'', j'')$ share a node\},
    \item $W(A_{(i, j)}) = \sum_{(i', j')\in A_{(i, j)}}w_{i'j'}$, and $W(B_{(i, j)}) = \sum_{(i'', j'')\in B_{(i, j)}}w_{i''j''}$.
\end{itemize}

\begin{theorem}\label{th:conver}
Under conditions: 
\begin{enumerate} 
\item[(i)] $G = \mathcal{O}(N^{\alpha}), 1\leq \alpha < 1.5$,
\item[(ii)] $S_1 + S_2 - \frac{4}{N}S_3 = \mathcal{O}(S_1 + S_2)$,
\item[(iii)] $\sum_{(i, j)\in G}(w_{ij}|A_{(i, j)}|)^2 = o(S_1\sqrt{N})$,
\item[(iv)] $\sum_{(i, j)\in G}w_{ij}W(A_{(i, j)})W(B_{(i, j)})= o(S_1)^{1.5}$, 
\end{enumerate} as $n_1$, $n_2$, $N\rightarrow\infty$ and $n_1/N \rightarrow \lambda \in (0, 1)$, $Z_w^R\xrightarrow{\mathcal{D}}N(0,1)$ and $Z_\text{diff}^R\xrightarrow{\mathcal{D}}N(0,1)$ under the permutation null distribution.
\end{theorem}
The proof utilizes Stein's theorem from \cite{chen2005stein}, and details are provided in the Appendix \ref{Appendix:limit_distribution}. 
\begin{corollary}
Under the conditions given in Theorem \ref{th:conver}, as $n_1, n_2, N\rightarrow\infty$ and $n_1/N \rightarrow \lambda \in (0, 1)$, $S_R\xrightarrow{\mathcal{D}}\chi_2$ under the permutation null distribution.
\end{corollary}

Condition (ii) ensures $Z^R_\text{diff}$ is asymptotically well-defined. The condition is automatically met when using the proposed weight function $W$ (\ref{eq:W}). Utilizing the proposed weight ensures $S_3$ is bounded by a constant independent of $N$. 

\begin{figure}[!ht]
\centering
\includegraphics[width=0.9\columnwidth]{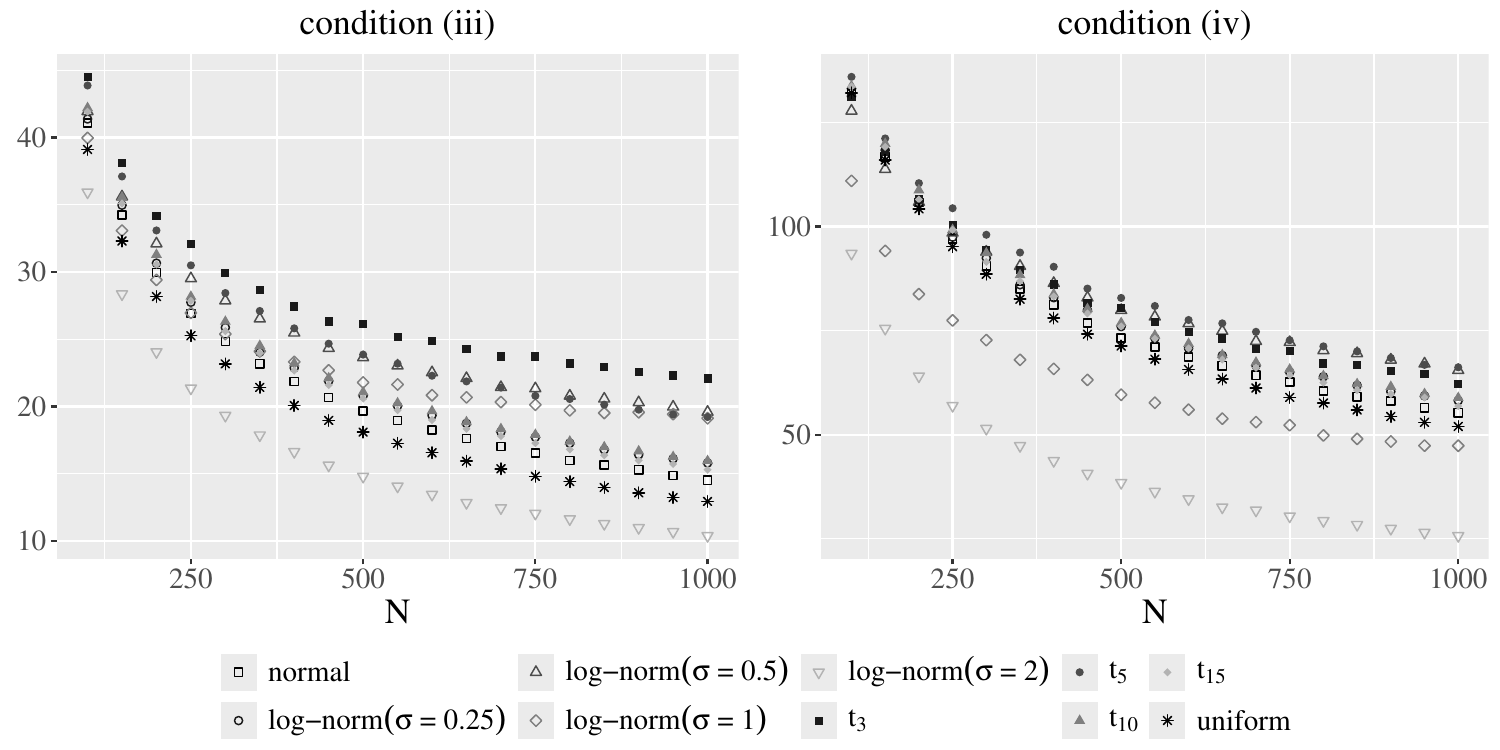}
\caption{Ratios of key quantities for the proposed robust test statistics generated using data from normal distribution, uniform distribution, log-normal distributions with different skewness levels (controlled by $\sigma$), and heavy-tailed $t$ distributions with varying degrees of freedom. The dimension of each observation is $d$ = $N$. Left: the ratio of $\sum(w_{ij}|A_{(i, j)}|)^2$ to $S_1N^{0.5}$. Right: the ratio of $\sum w_{ij}W(A_{(i, j)})W(B_{(i, j)})$ to $(S_1)^{1.5}$.}
\label{fig:add_lim1}
\end{figure}

Conditions (iii) and (iv) prevent the sum of weights in the hub from growing too large. To see that the conditions hold easily in the presence of hubs in high dimensions, we generated data from the normal distribution, uniform distribution, log-normal distributions, and heavy-tailed $t$ distributions. Ratios of the key quantities involved in the conditions are shown in Figure \ref{fig:add_lim1}. Once we assign weights, the ratios $\sum(w_{ij}|A_{(i, j)}|)^2/ (S_1\sqrt{N})$ and $(\sum w_{ij}W(A_{(i, j)})W(B_{(i, j)})/(S_1)^{1.5}$ are bounded by $o(1)$ as $N$ increases under all scenarios.

\begin{figure}[!ht]
\centering
\includegraphics[scale=0.65]{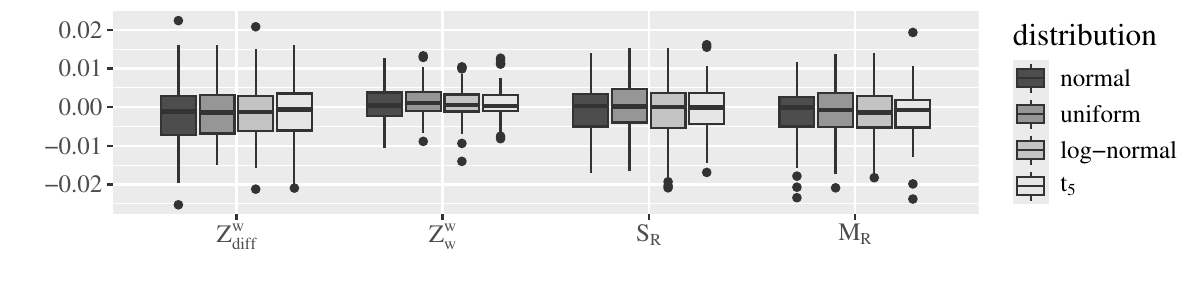}

\caption{Boxplots of differences between asymptotic critical values and permutation critical values. Data are generated from different distributions with $n_1 = n_2 = 100$ and $d = 100$.}
\label{fig:cri_value}
\end{figure}

To evaluate the accuracy of our asymptotic theory for finite sample sizes, we compare the critical values generated from 10,000 permutations with those obtained using our asymptotic theory under the null hypothesis. The boxplots of the differences between asymptotic and permutation critical values are shown in Figure \ref{fig:cri_value}. We observe that the p-value approximations are reasonable based on the small differences shown in the boxplots.

\section{Performance Analysis}\label{sec:performance}
\subsection{Hubs in high-dimensional data}

We examine the performance of the robust edge-count test statistics on high-dimensional data. We present the power of the tests, which is estimated to be the number of trials (out of 100) with significance less than 5\%. We also report the median of the maximum node degrees of $5$-MST in each trial (over 100 trials), denoted as $\tilde{d}_{\text{max}}$. We compare the robust tests $S_R$ and $M_R$ with the following tests: MMD (\cite{gretton2012kernel}), energy (\cite{szekely2004testing}), the generalized edge-count test $S$ (\cite{gen2017}), the max-type edge-count test $M$ (\cite{max2019}), and rank-based tests $R_g$-NN and $R_o$-MST (\cite{zhou2023rank}). The test statistics $R_g$-NN and $R_o$-MST also apply weights to a similarity graph (NN or MST) in the form of ranks. However, their ranking weights are not designed to mitigate problematic structures in the graph and, as we'll demonstrate, can still suffer from reduced power in some scenarios. For $R_g$-NN, we follow the authors' recommendation and use the 10-NN graph. The detailed settings of the simulations are as follows:
\begin{itemize}
\item Simulation I and III: Observations are generated from multivariate log-normal distributions. $$\textbf{X} \sim \text{exp}(\mathcal{N}(\textbf{1}_d, 0.6\textbf{I}_d)),$$ $$\textbf{Y} \sim \text{exp}(\mathcal{N}((1 + \sqrt{0.01log(d)/d})\textbf{1}_d, (0.6 + 1.8log(d)/d)\textbf{I}_d)),$$ where $d$ denotes the dimension. $n_1 = n_2 = 100$.
\item Simulation II and IV: Observations are generated from multivariate mixture Gaussian distributions. $$\textbf{X} \sim \mathcal{N}(\textbf{0}_d, \textbf{I}_d),$$ $$\textbf{Y} \sim 0.1\mathcal{N}(\textbf{0}_d, \textbf{I}_d) + 0.9\mathcal{N}(\sqrt{0.1log(d)/d}\textbf{1}_d, (1 + 2.5log(d)/d)\textbf{I}_d),$$ where $d$ denotes the dimension. $n_1 = n_2 = 100$. 
\item Simulation V: Observations are generated from multivariate Gaussian distributions: $$\textbf{X}:\mathcal{N}(\textbf{0}_{d}, \textbf{I}_{d}),$$ $$\textbf{Y}: \mathcal{N}(\boldsymbol{(\sqrt{(0.2log(s)/s)}1_{s}, 0_{d-s})}, \big(\begin{smallmatrix}
  (1+3log(s)/s)^2\textbf{I}_{s} & \textbf{0}\\
  \textbf{0} & \textbf{I}_{d-s}
\end{smallmatrix}\big)),$$
 where $s = \lfloor{\sqrt{d}}\rfloor.$ $n_1 = n_2 = 200$.
\end{itemize}

The results for Simulations I and II are presented in Table \ref{tab:prob-log-normal} and \ref{tab:prob-mixture}. In both settings, there is a mean and variance change. For each of the 100 trials, the maximum node degree appears in the sample with a larger variance. Both simulations exhibit a pattern of hubness in the high-dimensional setting, which creates difficulties for the existing two-sample tests. For log-normal data, the hubness is more pronounced as the dimension increases ($\tilde{d}_\text{max}$ is quite large). The MMD test struggles in this setting. When the dimension is moderate ($d=500$), the remaining non-parametric tests perform reasonably. However as $d$ increases, the power for some of the tests begins to suffer. We observe that as $d$ increases, the robust edge-count tests $S_R$ and $M_R$  outperform all other methods and the gap becomes more pronounced for larger $d$. We observe a similar pattern in Table \ref{tab:prob-mixture}: as $d$ increases the robust edge-count tests have considerable power gains compared to other methods. 

\begin{table*}[!ht]
    \caption{Simulation I: number of trials that reject the null with $\alpha =0.05$.}
    \begin{tabular*}{\columnwidth}{@{\extracolsep{\fill}}cccccccccc@{\extracolsep{\fill}}}
        \toprule
         $d$ &  $\tilde{d}_{\text{max}}$ & MMD& Energy& $S$  & $M$ & $R_g$-NN & $R_o$-MST & $S_R$ & $M_R$ \\
      
        \midrule
         500 & 111.5 & 6 & 75 & 55 & 66 & 67 & 89 & 97 & \textbf{98}\\
         800 & 128 & 4 & 51 & 40 & 48 & 46 & 81 & 89 & \textbf{93}\\
         1100 & 124 & 6 & 33 & 36 & 39 & 36 & 77 & \textbf{90} & \textbf{90}\\
         1400 & 131 & 4 & 19 & 20 & 28 & 18 & 65 & 87 & \textbf{90}\\
         1700 & 127.5 & 5 & 17 & 10 & 15 & 13 & 53 & 77 & \textbf{80}\\
         2000 & 134.5 & 7 & 14 & 15 & 24 & 18 & 54 & 70 & \textbf{76}\\
        \bottomrule
     \end{tabular*}
    \label{tab:prob-log-normal}
\end{table*}

\begin{table*}[!ht]
    \caption{Simulation II: number of trials that reject the null with $\alpha =0.05$.}
    \begin{tabular*}{\columnwidth}{@{\extracolsep{\fill}}cccccccccc@{\extracolsep{\fill}}}
        \toprule
         $d$ &  $\tilde{d}_{\text{max}}$ & MMD& Energy& $S$  & $M$ & $R_g$-NN & $R_o$-MST & $S_R$ & $M_R$ \\
      
        \midrule
         500 & 69 & 26 & 30 & 84 & 84 & 84 & 91 & \textbf{100} & \textbf{100}\\
         800 & 71 & 14 & 19 & 56 & 68 & 55 & 76 & 92 & \textbf{93}\\
         1100 & 74 & 8 & 20 & 53 & 49 & 52 & 62 & 84 & \textbf{91}\\
         1400 & 71 & 6 & 14 & 35 & 43 & 35 & 56 & 80 & \textbf{86}\\
         1700 & 71 & 8 & 15 & 27 & 31 & 31 & 50 & 66 & \textbf{73}\\
         2000 & 71 & 5 & 17 & 23 & 26 & 23 & 38 & 64 & \textbf{65}\\
        \bottomrule
     \end{tabular*}
    \label{tab:prob-mixture}
\end{table*}

The results for Simulations III and IV are presented in Table \ref{tab:log-normal} and \ref{tab:mixture}. These settings are similar to Simulations I and II but allow the maximum node degree to appear in either sample. Under these scenarios, the robust edge-count tests show comparable results to $R_o$-MST while outperforming other tests for the log-normal data (see Table \ref{tab:log-normal}). In Table \ref{tab:mixture}, we observe that the robust edge-count tests excel when compared to other tests for the mixture Gaussian data as the dimension of the observation increases.

\begin{table*}[!ht]
    \caption{Simulation III: number of trials that reject the null with $\alpha =0.05$.}
    \begin{tabular*}{\columnwidth}{@{\extracolsep{\fill}}cccccccccc@{\extracolsep{\fill}}}
        \toprule
         $d$ &  $\tilde{d}_{\text{max}}$ & MMD& Energy& $S$  & $M$ & $R_g$-NN & $R_o$-MST & $S_R$ & $M_R$ \\
      
        \midrule
         500 & 123 & 50 & 83 & 91 & 91 & 93 & \textbf{100} & 98 & 99\\
         800 & 130 & 44 & 61 & 77 & 83 & 81 & 96 & 95 & \textbf{97}\\
         1100 & 137 & 37 & 45 & 81 & 82 & 81 & 94 & 94 & \textbf{98}\\
         1400 & 134 & 35 & 29 & 71 & 78 & 70 & \textbf{94} & 89 & \textbf{94}\\
         1700 &  137.5 & 32 & 19 & 62 & 72 & 62 & \textbf{90} & 85 & \textbf{90}\\
         2000 &  143.5 & 26 & 19 & 50 & 63 & 54 & 85 & 80 & \textbf{86}\\
        \bottomrule
     \end{tabular*}
    \label{tab:log-normal}
\end{table*}

\begin{table*}[!ht]
    \caption{Simulation IV: number of trials that reject the null with $\alpha =0.05$.}
    \begin{tabular*}{\columnwidth}{@{\extracolsep{\fill}}cccccccccc@{\extracolsep{\fill}}}
        \toprule
         $d$ &  $\tilde{d}_{\text{max}}$ & MMD& Energy& $S$  & $M$ & $R_g$-NN & $R_o$-MST & $S_R$ & $M_R$ \\
      
        \midrule
         500 & 74 & 49 & 25 & 96 & \textbf{99} & 96 & \textbf{99} & 98 & 98\\
         800 & 74.5 & 38 & 20 & 90 & 92 & 89 & 96 & 99 & \textbf{100}\\
         1100 & 72.5 & 24 & 13 & 76 & 89 & 77 & 86 & 91 & \textbf{94}\\
         1400 & 75 & 25 & 14 & 71 & 75 & 70 & 83 & 86 & \textbf{90}\\
         1700 &  72 & 29 & 15 & 78 & 82 & 78 & 87 & 83 & \textbf{88}\\
         2000 &  72 & 17 & 11 & 53 & 55 & 47 & 65 & 73 & \textbf{77}\\
        \bottomrule
     \end{tabular*}
    \label{tab:mixture}
\end{table*}

Table \ref{tab:sparse} presents the performance under Simulation V. We simulate observations where the change does not occur in all dimensions; this setting can easily induce a large hub in the similarity graph when the dimension is high. Similar to before, when the dimension is not too high ($d=500$) all the tests have comparable power. But as $d$ increases, we see that the robust tests start to out-compete most of the other graph-based tests. When $d = 2000$, it is evident that the robust tests have the superior power.

\begin{table*}[!ht]
    \caption{Simulation V: number of trials that reject the null with $\alpha =0.05$.}
    \begin{tabular*}{\columnwidth}{@{\extracolsep{\fill}}cccccccccc@{\extracolsep{\fill}}}
        \toprule
         $d$& $\tilde{d}_{\text{max}}$ & MMD & Energy &$S$  & $M$ & $R_g$-NN & $R_o$-MST & $S_R$ & $M_R$ \\
      
        \midrule
         500 &  107& 47& 65& \textbf{100} & \textbf{100}  & \textbf{100} & \textbf{100} & \textbf{100} & \textbf{100} \\
         800 & 109& 41&61 &91 & 92 & 93 & \textbf{98} & 96 & 96\\
         1100 & 112& 34& 46 & 86 & 89 & 87 & 92 & \textbf{95} & \textbf{95} \\
         1400 &109& 34& 45 &75 & 79 & 78 & 87 & 87 & \textbf{92} \\
         1700 & 107& 21&42 &72 & 79 & 70 & 84 & 84 & \textbf{87} \\
         2000 & 108& 20& 33 &67 & 70 & 68 & 77 & 84 & \textbf{85} \\
        \bottomrule
     \end{tabular*}
    \label{tab:sparse}
\end{table*}

The robust tests are also well-designed to deal with the hubness phenomenon under imbalanced sample sizes as explored in previous studies \cite{wei2018, banerjee2020nearest, banerjee2023bootstrapped}. In particular, the test statistic $Z^R_w$ is constructed to mitigate any power loss from imbalanced samples.  Since $S_R$ and $M_R$ are functions of $Z^R_w$, both test statistics are equipped to handle the imbalanced setting and hubness phenomenon for general changes. Under the imbalanced setting, the larger sample is more likely to develop a hub.  Additional simulations demonstrating the performance of the robust graph-based tests under imbalanced sample sizes are provided in Appendix \ref{Appendix:imbalance}.

\subsection{Calibration Under the Null Hypothesis}\label{sec:sim_null}
To assess the calibration of robust edge-count tests under the null hypothesis, we simulate two samples with $n_1 = n_2 = 100$ from the standard normal distribution. In Table \ref{tab:null}, the number of trials (out of 1000) to reject the null (at $\alpha = 0.05$) are reported for both asymptotic and permutation critical values. Rejection rates are around 5\% for dimensions ranging from 600 to 2000, indicating that the type I error rate is well-controlled.

\begin{table*}[!ht]
    \caption{Number of trials (out of 1000) that reject the null with $\alpha = 0.05$ under the null hypothesis.}
    \begin{tabular*}{\columnwidth}{@{\extracolsep{\fill}}c|ccccccccc@{\extracolsep{\fill}}}
        \toprule
        
&$d$& 600 & 800 & 1000 & 1200 & 1400 & 1600 & 1800 & 2000\\
\midrule
\multirow{2}{*}{Permutation}&$S_R$ & 45 & 48 & 53& 42& 42& 38 & 55 & 49\\
&$M_R$ & 44 & 43 & 62& 53& 46& 45 & 51 & 46 \\
        \midrule
\multirow{2}{*}{Asymptotic}&$S_R$ & 39 & 45  & 51& 41& 41& 38 & 57 & 48\\
&$M_R$ & 45 & 42 & 59& 50& 46& 44 & 50 & 46\\
        \bottomrule
     \end{tabular*}
    \label{tab:null}
\end{table*}

\subsection{Consistency of the proposed tests\label{sec:sim_consis}}

The robust edge-count tests show increasing power as the number of observations grows. We simulate samples with various sample sizes to exhibit the consistency of the test. The simulated data are generated from log-normal distributions with
$\textbf{X} \sim \text{exp}(\mathcal{N}(\textbf{1}_d, 0.6\textbf{I}_d))$, $\textbf{Y} \sim \text{exp}(\mathcal{N}((1 + \sqrt{0.01log(d)/d})\textbf{1}_d, (0.6 + 1.8log(d)/d)\textbf{I}_d))$, and mixture Gaussian distributions with 
 $\textbf{X} \sim \mathcal{N}(\textbf{0}_d, \textbf{I}_d)$, $\textbf{Y} \sim \lfloor0.1n_2\rfloor\mathcal{N}(\textbf{0}_d, \textbf{I}_d) + \lfloor0.9n_2\rfloor$$\mathcal{N}(\sqrt{0.1log(d)/d}\textbf{1}_d,$\\$ (1 + 2.5log(d)/d)\textbf{I}_d)$, where $d = 2000$. 
The powers of the robust edge-count tests at 5\% significance level for 100 simulations are presented in Table \ref{tab:consis}. With more observations, the number of rejections increases and quickly approaches 100, even for moderate sample sizes, demonstrating the consistency of the proposed tests.

Following Theorem 5.2.1 from \cite{gen2017}, it is straightforward to show that the robust edge-count tests are consistent against all alternatives on $k$-MST with $k = O(1)$.% by incorporating weight into the asymptotic distribution.
\begin{table*}[!ht]
    \caption{Number of trials (out of 100) that reject the null for $\alpha = 0.05$ as $N = n_1 + n_2$ increases.}
    \begin{tabular*}{\columnwidth}{@{\extracolsep{\fill}}ccccccccccc@{\extracolsep{\fill}}}
        \toprule
         $n_1 = n_2$&&100 & 125 & 150 &175& 200 & 225 & 250 & 275 & 300 \\
      
        \midrule
         \multirow{2}{*}{log-normal} & $S_R$ &76& 76& 93& 93 & 94 & 97 &  100& 100 &  100\\
         & $M_R$ &85& 84&95 & 93 &  98 & 97 & 100 & 100 & 100 \\
         \multirow{2}{*}{mixture Gaussian}  & $S_R$ &83 &92 &91 &99 & 99 & 99 & 99 & 100 &100 \\
         & $M_R$ &85 &94 & 90&99 & 100 & 99 & 99 & 100 & 100\\
        \bottomrule
     \end{tabular*}
    \label{tab:consis}
\end{table*}

\section{Real Data Application}\label{sec:application}
\begin{figure}[!ht]
\centering
\subfloat[Pick-up locations in Chicago.]{\label{fig:location}
\centering
\includegraphics[width=0.46\linewidth]{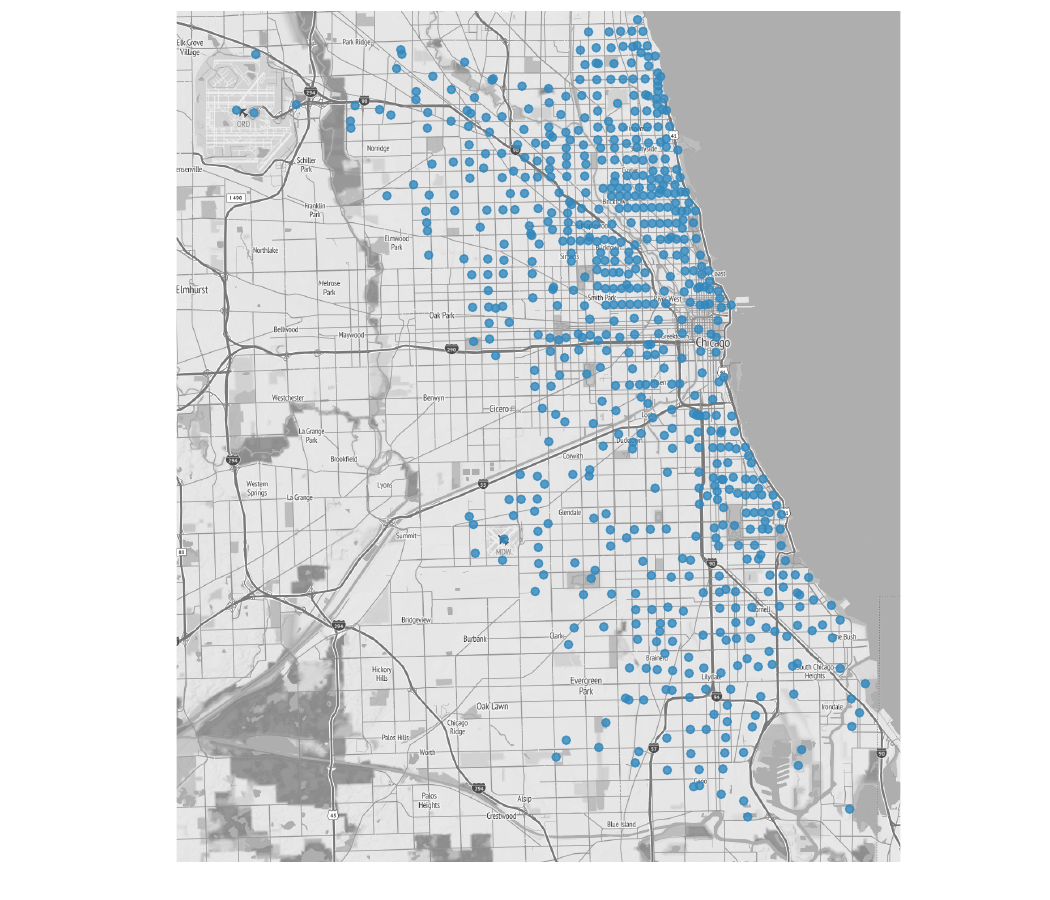}
}
\subfloat[Frequency of taxi dropoffs.]{\label{fig:one_unit}
\centering
\includegraphics[width=0.46\linewidth]{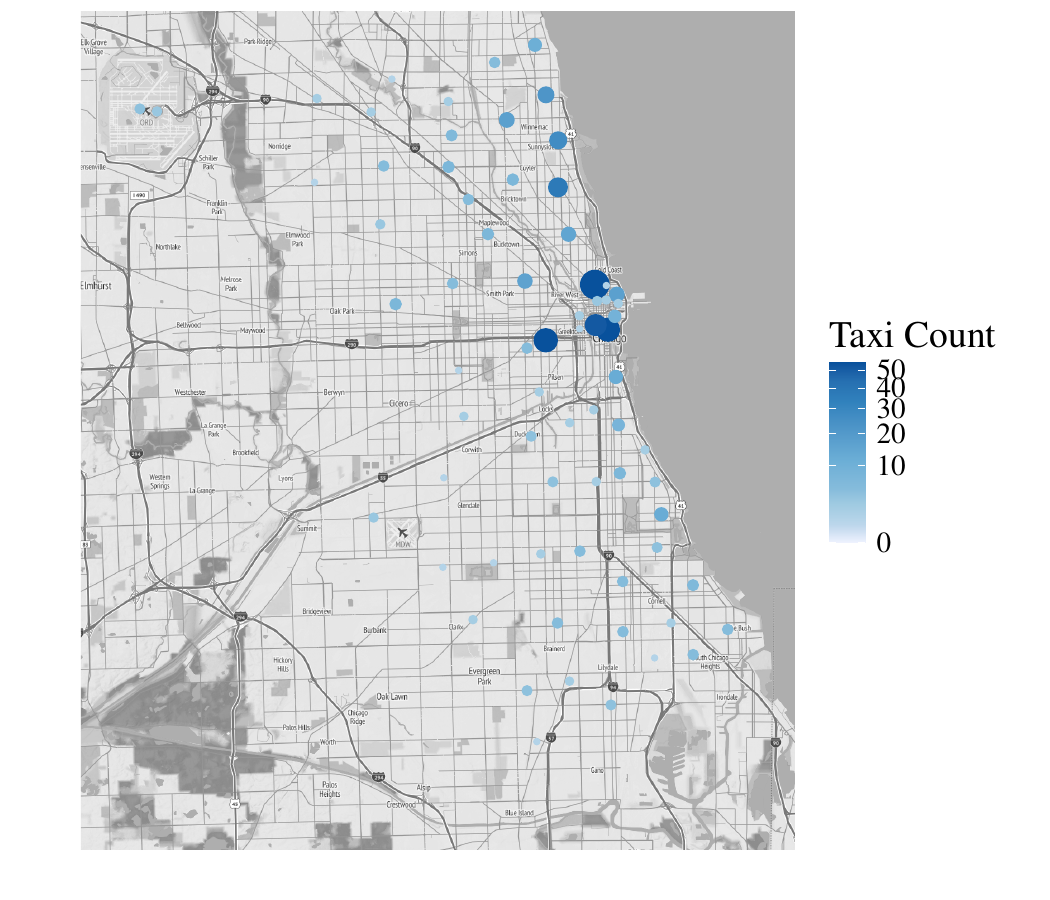}
}
\caption{Left: plot of all drop-off locations. Right: plot of the number of taxi trips that occurred from 7 am to 10 am on September 1st at all possible drop-off locations. The larger dot size indicates more trips took place at the location. A color key is also provided.}
\label{fig:chicago}
\end{figure}

We illustrate the robust graph-based tests on the Chicago taxi trip dataset in 2020. This data is publicly available on the Chicago Data Portal website  (\url{https://data.cityofchicago.org/Transportation/Taxi-Trips/wrvz-psew}) and includes drop-off dates, times, and locations for each taxi trip. There are 635 unique drop-off locations (as shown in Figure \ref{fig:location}). We count the frequency of taxi drop-offs in each location for a specified time interval. Each observation is a $635\times 1$ vector of taxi trip counts that occur within a time interval for each day; each element of the vector represents the number of drop-offs at a specific drop-off location. In Figure \ref{fig:location}, the position of the dot indicates the location of taxi drop-offs. Figure \ref{fig:one_unit} shows an example when the time interval is set to be 7 am - 10 am on September 1.  The size and color of the dot indicate the number of trips at that location.

Since taxi trips may not happen in some locations for a specified time interval, it is often the case that many entries in our vector of observations are 0 or very low counts. Given that the dimension of the observation is much larger than the number of observations in each sample, a large node degree is likely to arise when constructing the similarity graph. Since the underlying data distribution is unknown, it is difficult to identify problematic hubs just by examining the constructed similarity graph. We will demonstrate that the robust edge-count test can circumvent any hub-related issues and lead to reasonable inference. 

To illustrate the new tests, we consider two different scenarios and compare the performance of the new tests with existing graph-based tests, as well as the energy test and MMD test. For similarity graphs, we use $5$-MST constructed from the $L_1$ distance between observations. For all tests, p-values are obtained via 10,000 permutations. 

\subsection{Scenario I} \label{sec:app0}
\begin{figure}[t!]
\centering 
\includegraphics[scale=0.4]{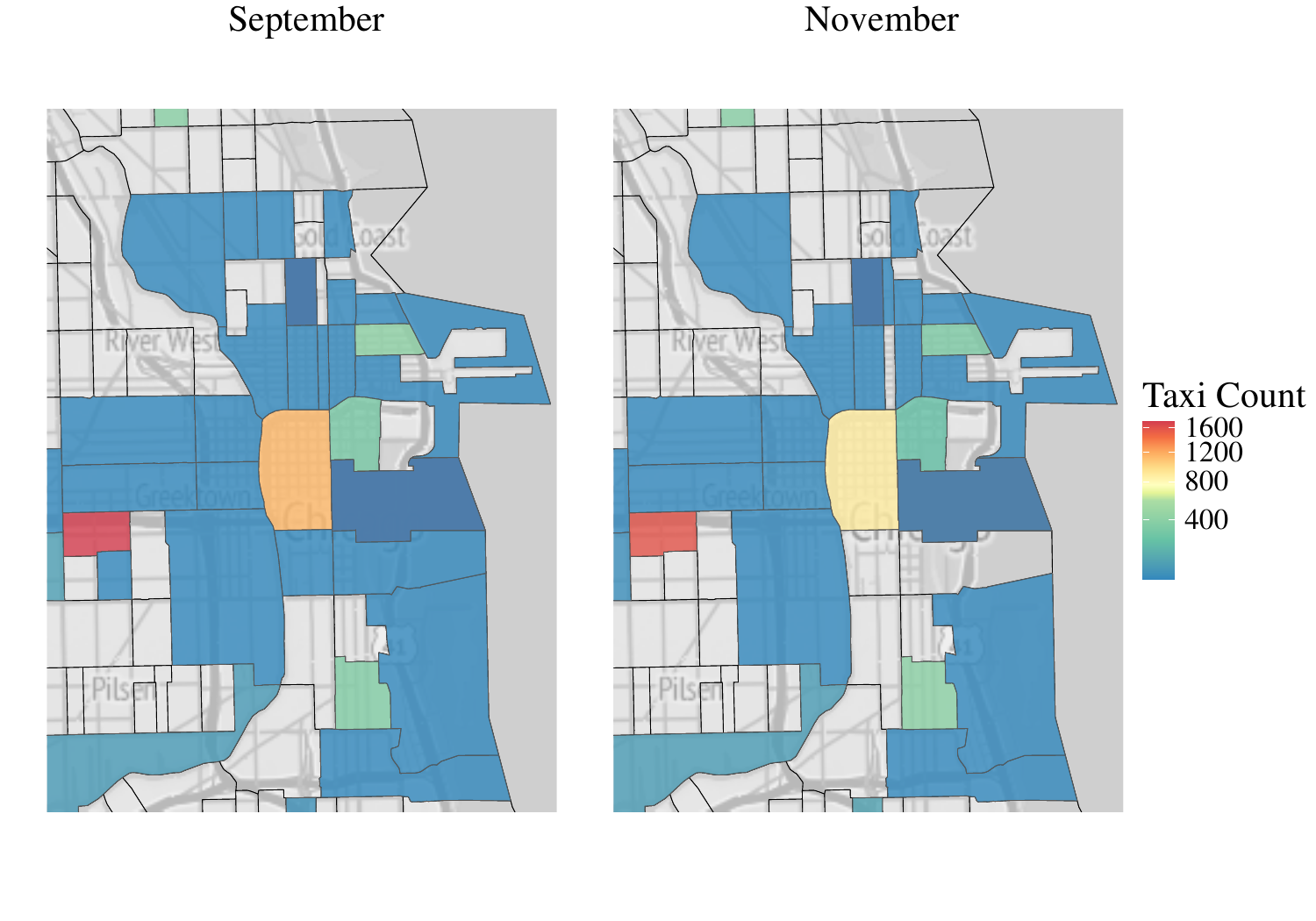}\par
\caption{Heatmap illustrating the number of taxi trips in each district for the month of September (left) and November (right).}
\label{fig:SeptNov}
\end{figure}

We compare the taxi drop-offs in morning rush hours from 7 am to 10 am between September and November. Sample 1 consists of the number of taxi drop-offs that occurred during morning rush hour in September. Each day is an observation, resulting in 30 observations ($n_1 = 30$). Sample 2 consists of the number of taxi drop-offs that occurred in the morning hours in November, with each day being an observation ($n_2 =30$). The dimension of each observation is 635, which is clearly far more than the number of observations. The heatmaps of the taxi counts in each district are shown in Figure \ref{fig:SeptNov}. The changes are subtle but taxi trips in September appear busier and more dispersed than in November. While we might be able to visualize this difference between months, what we want to effectively discern is whether this change in distribution is meaningful or just by random chance. 

To address this question, Table \ref{tab:sepnov} presents the two-sample test results. With a max node degree of 19 in the similarity graph, all graph-based tests provide significant evidence in favor of a difference between September and November at a 10\% significance level. However, the energy and MMD tests cannot reject the null hypothesis. This demonstrates, at least in this setting, that the graph-based methods show superior performance as omnibus tests when comparing high-dimensional distributions. 

\begin{table*}[!ht]
 \caption{P-values for tests comparing taxi dropoffs in morning rush hours between September and November.}
 \begin{tabular*}{\columnwidth}{@{\extracolsep{\fill}}cccccccc@{\extracolsep{\fill}}}
        \toprule
MMD & Energy & $S$ &$M$ &$S_R$ &$M_R$ \\
        \midrule
  0.3871 & 0.1696 & 0.0032 & 0.0024 & 0.0024 & 0.0025\\
        \bottomrule
    \end{tabular*}
       \label{tab:sepnov}
\end{table*}

As a sanity check, to demonstrate that the tests are well-calibrated, we randomly split the morning rush hour taxi dropoffs in September and November into two samples. As shown in table \ref{tab:sepnov_ran}, all tests fail to reject the null hypothesis of no difference at a 10\% significance level.

\begin{table*}[!ht]
 \caption{P-values for tests comparing randomly split taxi drop-offs in morning rush hours in September and November.}
\begin{tabular*}{\columnwidth}{@{\extracolsep{\fill}}cccccccc@{\extracolsep{\fill}}}
        \toprule
MMD & Energy & $S$ &$M$ &$S_R$ &$M_R$ \\
        \midrule
  0.9293 & 0.5720 & 0.9933 & 0.9217 & 0.9912 & 0.9630\\
        \bottomrule
    \end{tabular*}
       \label{tab:sepnov_ran}
\end{table*}

\subsection{Scenario II}\label{sec:app1}
The statewide stay-at-home order signed by the Illinois Governor took effect on March 21 in response to the spread of COVID-19, leading to a sharp decline in Chicago taxi trips. This is a setting where large hubs cause issues for the existing graph-based tests. In the early morning hours (1 am - 5 am), the number of taxi rides is relatively low, especially after the lockdown in March; this sparse taxi activity induces the formation of hubs with large node degrees in the high-dimensional setting. We compare the number of taxi dropoffs during the early morning hours between weekdays and weekends over two months (April and May). Sample 1 comprises the taxi dropoffs on 43 weekdays during early morning hours $(n_1 = 43)$, while Sample 2 comprises the taxi dropoffs on 18 weekends during early morning hours $(n_2 = 18)$. 

\begin{figure}[ht!]
\centering
\includegraphics[width=0.6\columnwidth]{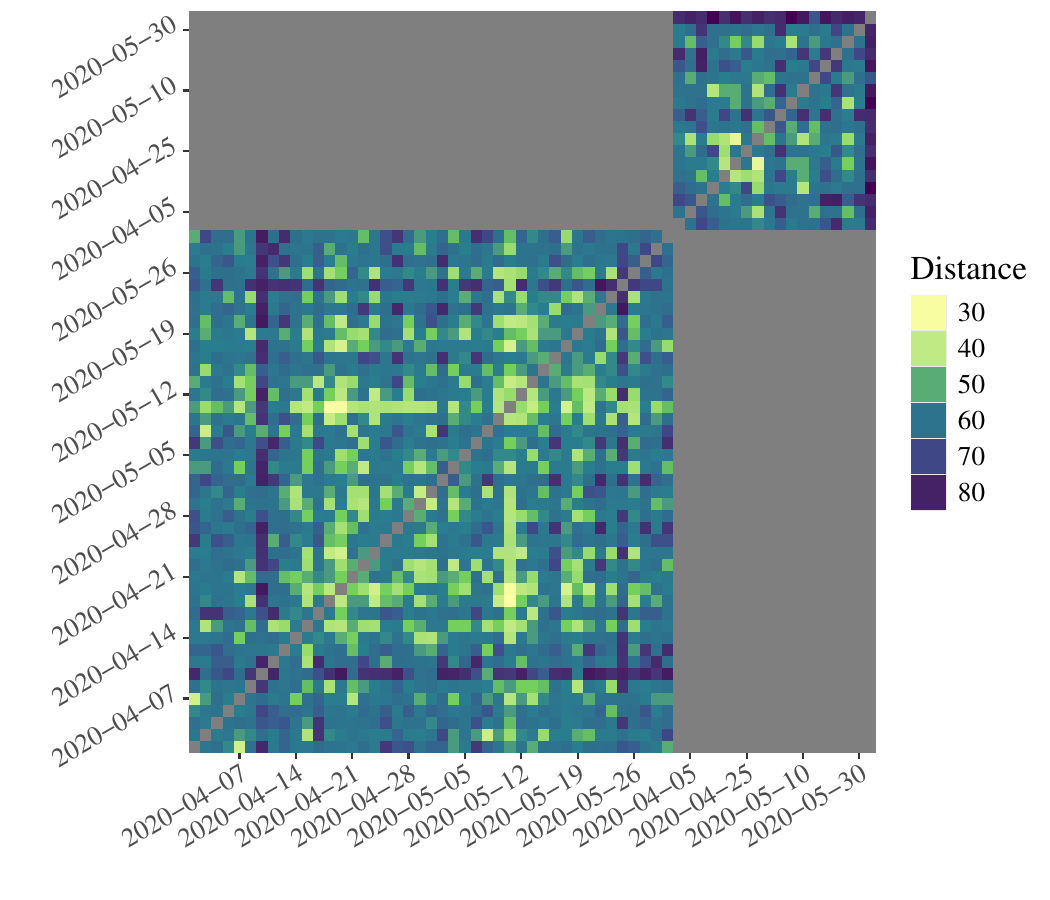}
\caption{Heatmap of the pairwise distance within the two samples (weekdays versus weekends). The bottom left square shows the pairwise distance for the weekday taxi dropoffs, and the top right square for weekends. Lighter colors indicate closer distances. For ease of visualization, the between-sample distances are not shown.}
\label{fig:early_AprilMay}
\end{figure}

Figure \ref{fig:early_AprilMay} displays the heatmap of the pairwise distance within two samples, indicating that the weekday observations tend to be closer (more similar) compared to weekends. We conjecture that there is a difference between the weekdays and weekends in the early morning, even post-lockdown. The similarity graph generated has several hubs with node degrees exceeding 20. Table \ref{tab:early_AprilMay} presents the two-sample test results. In this scenario, only the robust tests $S_R$ and $M_R$ provide evidence of a difference in travel patterns between the two samples at $\alpha = 0.1$. The MMD and energy test, while not significant, have p-values that are seemingly in the right direction compared to the other graph-based tests $S$ and $M$.  

\begin{table*}[ht]
 \caption{P-values for tests comparing taxi dropoffs on weekdays and weekends in April and May.}
 \begin{tabular*}{\columnwidth}{@{\extracolsep{\fill}}cccccccc@{\extracolsep{\fill}}}
        \toprule
MMD & Energy & $S$ &$M$  &$S_R$ &$M_R$ \\
        \midrule
       0.1289 & 0.1098 & 0.2617 & 0.2150 & 0.0223 & 0.0804\\
        \bottomrule
    \end{tabular*}
       \label{tab:early_AprilMay}
\end{table*}

To better understand the behavior of the graph-based tests, we conduct a small sensitivity analysis to see how observations with large node degrees influence the tests' conclusions. One influential observation generating a hub with node degree of 31 is from April 26. As shown in Table \ref{tab:detail_AprilMay}, after removing this observation, % taxi trips on April 24, 
the MMD test still cannot reject the null, while the other tests have significant test results rejecting the null at $\alpha = 0.1$. 

\begin{table*}[ht]
 \caption{P-values for tests comparing taxi dropoffs on weekdays and weekends in April and May after removing activity on April 26th.}
 \begin{tabular*}{\columnwidth}{@{\extracolsep{\fill}}cccccccc@{\extracolsep{\fill}}}
        \toprule
MMD & Energy & $S$ &$M$  &$S_R$ &$M_R$ \\
        \midrule
       0.1217 & 0.0977 & 0.0432 & 0.0774 & 0.0149 & 0.0600 \\
        \bottomrule
    \end{tabular*}
       \label{tab:detail_AprilMay}
\end{table*}

Table \ref{tab:wdwe_ran} shows that for this setting, the tests are well-calibrated under the null. By randomly assigning the early morning hour taxi drop-offs in April and May into Sample 1 ($n_1 = 43$) and Sample 2 ($n_2 = 18$), all tests fail to reject the null at a 10\% significance level.

\begin{table*}[!ht]
 \caption{P-values for tests comparing randomly split taxi dropoffs in the early morning hours on weekdays and on weekends in April and May.}
 \begin{tabular*}{\columnwidth}{@{\extracolsep{\fill}}cccccccc@{\extracolsep{\fill}}}
        \toprule
MMD & Energy & $S$ &$M$ &$S_R$ &$M_R$ \\
        \midrule
  0.9144 & 0.9467 & 0.3846 & 0.3587 & 0.5849 & 0.4393\\
        \bottomrule
    \end{tabular*}
       \label{tab:wdwe_ran}
\end{table*}

In practice, identifying the influential observations, (such as the taxi drop-offs on April 26th), is challenging. The problematic observation may not necessarily be largest hub (node with max degree) but a collection of hubs. The influence of an observation depends heavily on the connectivity of the graph and which edges are connected to this hub. While the inclusion and exclusion of potentially problematic observations may lead to conflicting results across existing tests, in contrast, the proposed robust test statistics ($S_R$ and $M_R$) are shown to provide consistent and stable results. This is crucial for drawing statistical conclusions in real data applications where the ground truth is unknown.

\section{Discussion}\label{sec:conclusion}
In this article, we propose robust edge-count two-sample tests that provide reliable inference even in the presence of problematic graph structures that arise as a product of the curse of dimensionality. Our proposed robust tests can outperform the existing non-parametric tests in the presence of the hubs while providing comparable power even when the graph is relatively flat. The robust edge-count tests are constructed by applying weights that are functions of node degrees to the edge counts. A specific weight function with desirable properties is recommended.

These robust test statistics are computationally straightforward to calculate. While finite-sample $p$-values can be obtained via permutations, to make the test more computationally tractable, the limiting null distributions of the robust test statistics are derived under some mild conditions on the data-driven weights. Through empirical studies, these conditions are shown to be easily satisfied even in the presence of hubs. The $p$-value approximations based on asymptotic results are reasonably close to the permutation $p$-value for finite sample sizes, making the approach easy to apply to large data sets when permutation may be computationally prohibitive. Simulation studies show that the robust edge-count tests have power gains over existing edge-count tests when the dimension increases and hubs are more easily generated. An application of the tests on Chicago taxi data demonstrates the robust test statistics utility in high-dimensional settings. 

Our results pave the way for future work in a few directions. It is of great interest to obtain optimal weights for robust graph-based test statistics. While this may be difficult to derive for generic similarity graphs, we may first focus on well-behaved graphs such as $k$-MSTs or $k$-NNs. Secondly, the study could be broadened to incorporate dense similarity graphs (where $k = O(n^\alpha), 0 \le \alpha < 1$), which would require more careful theoretical analysis. Lastly, the robust edge-count tests can be adapted to the scan statistic setting and applied to high-dimensional change-point problems, where the effect of the hubs over time may hamper our ability to effectively detect changes in distribution.

%%%%%%%%%%%%%%

\bibliographystyle{amsplain}
\bibliography{bibliography}

\appendix

\section{Additional Simulations} 

\subsection{Simulation results of SHP test and cross-match test}
\label{Appendix:con_graph}

 In this section, we explore the performance of the Shortest Hamiltonian path (SHP)-based test \cite{biswas2014distribution} and the cross-match test based on non-bipartite matching \cite{rosenbaum2005exact} in the high-dimensional setting. 

Observations are simulated under distributional changes. Specifically, the simulation settings are as follows:
        \begin{itemize}
            \item Mean change only. Observations are generated from multivariate normal distributions: $X \sim \mathcal{N} (1_d , I_d )$, $Y \sim \mathcal{N} (\sqrt{1.5log(d)/d)}1_d, I_d)$, where $d$ denotes the dimension. $n_1 = n_2 = 100$.
            \item Scale change only. Observations are generated from multivariate normal distributions: $X \sim \mathcal{N} (1_d , I_d )$, $Y \sim \mathcal{N} (1_d, (1+1.5log(d)/d)I_d)$, where $d$ denotes the dimension. $n_1 = n_2 = 100$.
        \end{itemize}
        The SHP-based test and the cross-match test are designed using a similar rationale as the original graph-based test proposed by Friedman and Rafsky \cite{ori1979}. As such, these tests focus on the between-sample edge counts in the test statistic, which can encounter problems detecting general changes as the dimension $d$ increases \cite{gen2017}.
        We compare their performances to the robust edge-count tests $S_R$ and $M_R$ (introduced in Section \ref{sec:method} in the paper). From Table \ref{tab:shpmdp}, we can see the SHP-based test and cross-match test have reasonable power when $d = 500$ and $d=800$ for mean change, but its power starts to decay as $d$ increases. Under scale change, both have lower power than the robust edge-count tests; the cross-match test in particular seems to struggle in this setting. As $d$ goes to 2000, both robust edge-count tests demonstrate superior power.
        
\begin{table*}[!ht]
    \caption{Number of trials with significance less than 5\% for comparison of robust graph-based test $S_R$, $M_R$, SHP-based test and cross-match test with mean change and scale change.}
    \begin{tabular*}{\columnwidth}{@{\extracolsep{\fill}}c|cccc|cccc@{\extracolsep{\fill}}}
        \toprule
        & \multicolumn{4}{c|}{mean change} &\multicolumn{4}{c}{scale change}\\
        \midrule
         $d$ & SHP& cross-match& $S_R$ & $M_R$ & SHP& cross-match& $S_R$ & $M_R$ \\
      
        \midrule
         500 & 95 & 83 & \textbf{100} & \textbf{100} & 76 & 37 & \textbf{100} & \textbf{100} \\
         800 & 92 & 84 & 98 & \textbf{100} & 67 & 24 & \textbf{99} & \textbf{99}\\
         1100 & 77 & 67 & 95 & \textbf{97} & 55 & 20 & \textbf{97} & 95 \\
         1400 & 68 & 62 & \textbf{93} & 92 & 43 & 15 & 94 & \textbf{97} \\
         1700 &  66 & 57 & 91 & \textbf{92} & 35 & 16 & \textbf{93} & 92\\
         2000 &  71 & 55 & 92 & \textbf{96} & 35 & 24 & \textbf{88} & 86 \\
        \bottomrule
     \end{tabular*}
    \label{tab:shpmdp}
\end{table*}

\subsection{Simulation results of robust edge-count tests under imbalanced sample sizes}\label{Appendix:imbalance}

We carry out simulations to demonstrate the performance of the tests under imbalanced sample sizes. The data are simulated using the same settings as those in Simulation III in Section \ref{sec:performance}:
     $$\textbf{X} \sim \text{exp}(\mathcal{N}(\textbf{1}_d, 0.6\textbf{I}_d))$$
     $$\textbf{Y} \sim \text{exp}(\mathcal{N}((1 + \sqrt{0.01log(d)/d})\textbf{1}_d, (0.6 + 1.8log(d)/d)\textbf{I}_d)),$$
     where $d$ denotes the dimension. We investigate two unbalanced settings with different sample sizes of the two samples. As shown in Table \ref{tab:unb2} and \ref{tab:unb4}, the robust edge-count tests $S_R$ and $M_R$ still retain good performance across all imbalanced settings, and demonstrate improvement compared to the edge-count tests $S$ and $M$. When the sample sizes are not too unbalanced (Table \ref{tab:unb2}), most of the graph-based tests are on equal footing. However, when the imbalance between samples becomes more severe (Table \ref{tab:unb4}), all tests have diminished power. We observe that the hubness phenomenon is not exacerbated by the imbalanced sample size - both settings have max node degrees of similar sizes (142 and 138, when $d=2000$, respectively). However, hubness is still clearly a  problem here, since the new proposed tests tend to have better (or comparable) power across all settings. When the sample sizes are severely unbalanced (Table \ref{tab:unb4}), we see the new proposed robust tests are still performing quite well.   

   \begin{table*}[!ht]
    \caption{Number of trials with significance less than 5\%. $n_1 = 50$, $n_2 = 150$.}
    \begin{tabular*}{\columnwidth}{@{\extracolsep{\fill}}cccccccccc@{\extracolsep{\fill}}}
        \toprule
         $d$ &  $\tilde{d}_{\text{max}}$ & MMD& Energy& $S$  & $M$ & $R_g$-NN & $R_o$-MST & $S_R$ & $M_R$ \\
      
        \midrule
         500 & 124 & 52 & 26 & 96 & 96 & 97 & 97 & \textbf{99} & \textbf{99}\\
         800 & 130 & 49 & 11 & 90 & 90 & 90 & \textbf{97} & \textbf{97} & \textbf{97}\\
         1100 & 132 & 35 & 8 & 81 & 78 & 80 & 86 & 92 & \textbf{95}\\
         1400 & 137 & 35 & 3 & 66 & 78 & 70 & 87 & 90 & \textbf{91}\\
         1700 & 138 & 36 & 2 & 69 & 77 & 69 & \textbf{82} & 80 & \textbf{82}\\
         2000 & 142 & 31 & 6 & 71 & 68 & 72 & \textbf{83} & 81 & \textbf{83}\\
        \bottomrule
     \end{tabular*}
    \label{tab:unb2}
\end{table*}

\begin{table*}[!ht]
    \caption{Number of trials with significance less than 5\%. $n_1 = 15$, $n_2 = 185$.}
    \begin{tabular*}{\columnwidth}{@{\extracolsep{\fill}}cccccccccc@{\extracolsep{\fill}}}
        \toprule
         $d$ &  $\tilde{d}_{\text{max}}$ & MMD& Energy& $S$  & $M$ & $R_g$-NN & $R_o$-MST & $S_R$ & $M_R$ \\
      
        \midrule
         500 & 128 & 24 & 0 & 70 & 74 & 69 & 77 & \textbf{80} & \textbf{80}\\
         800 & 133 & 20 & 0 & 59 & 58 & 64 & 64 & \textbf{68} & 64\\
         1100 & 132 & 18 & 0 & 53 & 52 & 51 & 54 & 54 & \textbf{56}\\
         1400 & 138 & 15 & 1 & 47 & 48 & 52 & 52 & \textbf{56} & 53\\
         1700 & 140 & 18 & 0 & 39 & 41 & 38 & 41 & \textbf{47} & 46\\
         2000 & 138 & 16 & 0 & 43 & 44 & 43 & 51 & 48 & \textbf{53}\\
        \bottomrule
     \end{tabular*}
    \label{tab:unb4}
\end{table*}
\section{Extra Figures}\label{Appendix:extr_fig}

\subsection{Hubness phenomenon in high-dimensional data using $5$-NN}\label{Appendix:max_nodedegree_nn}

\begin{figure}[!t]
\centering
\includegraphics[width=\linewidth]{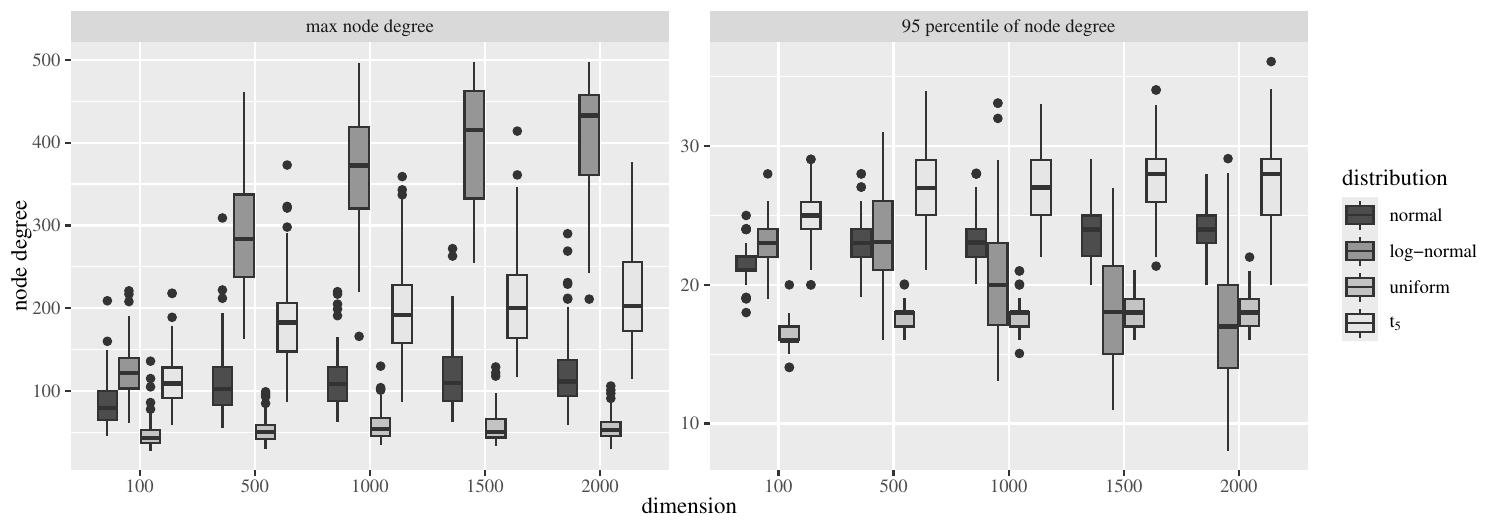}
\caption{Boxplot of maximum and 95th percentiles of node degrees for different dimensions. Results are from 100 simulations with $n = 500$, where observations are drawn from multivariate normal, log-normal, uniform, and t distributions.}
\label{fig:max_nodedegree_nn}
\end{figure}

The maximum and 95th percentile of node degrees in the similarity graph constructed using $5$-NN are shown in Figure \ref{fig:max_nodedegree_nn}. The hubness phenomenon is similar to what we can see using the $5$-MST as the similarity graph. The maximum node degrees are over three times as much as the 95th percentiles.

\section{Proof of Lemma \ref{th:meanvarori}} \label{Appendix:mean_var}
The mean and variance of $R_1^w$ under the permutation null distribution can be derived as follows:
\begin{equation*}
\begin{split}
\mu_1^w =& \sum_{(i, j) \in G} w_{ij}P(J_{(i, j)} = 1) 
	= \sum_{(i, j) \in G}w_{ij} \frac{n_1(n_1-1)}{N(N-1)},\\
    E((R_1^w)^2) =&  \sum_{(i, j), (k, l) \in G} w_{ij}w_{kl}P(J_{(i, j)} = 1, J_{(k, l)} = 1) \\ =& S_1 \frac{n_1(n_1-1)}{N(N-1)}+S'_2 \frac{n_1(n_1-1)(n_1-2)}{N(N-1)(N-2)}+\\
    &S'_3\frac{n_1(n_1-1)(n_1-2)(n_1-3)}{N(N-1)(N-2)(N-3)},\\
    \Sigma_{11} =& E((R_1^w)^2) - E^2(R_1^w),
\end{split}
\end{equation*}
where $
S_1 = \sum_{(i, j) \in G}w_{ij}^2$, $
S'_2 = \sum_{\substack{(i, j), (i, k) \in G\\ k, l \text{ are different}}}w_{ij}w_{ik}$, and \\ $
S'_3 = \sum_{\substack{(i, j), (k, l) \in G\\ i, j, k, l \text{ all different}}} w_{ij}w_{kl}.$

Similarly, we can get the mean and variance of $R_2^w$ under the permutation null distribution:
\begin{equation*}
\begin{split}
\mu_2^w =&  \sum_{(i, j) \in G} w_{ij}P(J_{(i, j)} = 2)  = \sum_{(i, j) \in G}w_{ij} \frac{n_2(n_2-1)}{N(N-1)},\\
 E((R_2^w)^2) =& S_1 \frac{n_2(n_2-1)}{N(N-1)}+S'_2 \frac{n_2(n_2-1)(n_2-2)}{N(N-1)(N-2)}+\\
 &S'_3\frac{n_2(n_2-1)(n_2-2)(n_2-3)}{N(N-1)(N-2)(N-3)}\\
    \Sigma_{22} =& E((R_2^w)^2) - E^2(R_2^w).
\end{split}
\end{equation*}

The covariance of $R_1^w$ and $R_2^w$ under the permutation null distribution can be derived as follows:
\begin{equation*}
\begin{split}
    E(R_1^wR_2^w) &=  \sum\limits_{(i, j), (k, l) \in G} w_{ij}w_{kl}P(J_{(i, j)} = 1, J_{(k, l)} = 2) \\& = S'_3\frac{n_1(n_1-1)n_2(n_2-1)}{N(N-1)(N-2)(N-3)},\\
    \Sigma_{12} &= E(R_1^wR_2^w) - E(R_1^w)E(R_2^w).
\end{split}
\end{equation*}

Note :
\begin{equation*}
\begin{split}
\sum_{\substack{{(i, j), (k, l) \in G}\\ {i, j, k, l \text{ all different}}}} w_{ij}w_{kl} = &\sum_{(i, j), (k, l) \in G} w_{ij}w_{kl} - \sum_{\substack{{(i, j), (i, k) \in G}\\ {k, l \text{ are different}}}} w_{ij}w_{ik} - \sum_{(i, j) \in G}w_{ij}^2,\\
\sum\limits_{\substack{{(i, j), (i, k) \in G}\\ {k, l \text{ are different}}}} w_{ij}w_{ik}=&\sum\limits_{(i, j), (i, k) \in G} w_{ij}w_{ik} -  \sum\limits_{(i, j) \in G}w_{ij}^2.\\
\end{split}
\end{equation*}

The variance and covariance can be simplified as
\begin{equation*}
\begin{split}
 \Sigma_{11} &= D_N \left\{\frac{N-3}{n_2-1}S_1+\frac{n_1-2}{n_2-1}S_2 +\frac{6(n_2-1)-4n_1(N-3)}{N(N-1)(n_2-1)}S_3\right\}\\
  &=D_N \left\{-S_2+\frac{2(2N-3)}{N(N-1)}S_3+ \frac{N-3}{n_2-1}   \left(S_1+S_2\right)-\frac{4(N-3)}{N(n_2-1)}S_3\right\},\\
  \Sigma_{12} &= D_N \left\{-S_2+\frac{2(2N-3)}{N(N-1)}S_3\right\},\\
\Sigma_{22}& =D_N \left\{\frac{N-3}{n_1-1}S_1+\frac{n_2-2}{n_1-1}S_2 +\frac{6(n_1-1)-4n_2(N-3)}{N(N-1)(n_1-1)}S_3 \right\}\\
  &=D_N \left\{-S_2+\frac{2(2N-3)}{N(N-1)}S_3+ \frac{N-3}{n_1-1}\left(S_1+S_2\right)-\frac{4(N-3)}{N(n_1-1)}S_3\right\},
\end{split}
\end{equation*}
where $S_1 = \sum_{(i, j) \in G}w_{ij}^2$, $S_2 = \sum_{(i, j), (i, k) \in G} w_{ij}w_{ik}$, $S_3 = \sum_{(i, j), (k, l) \in G} w_{ij}w_{kl}$ and $D_N = [n_1n_2(n_1-1)(n_2-1)]/[N(N-1)(N-2)(N-3)]$..
\section{Proof of Remark \ref{rm:welldefine}} \label{Appendix:welldefine}
\begin{equation*}
 \begin{split}
 \sum\limits_{(i, j) \in G}w_{ij}^2+\sum\limits_{(i, j), (i, k) \in G} w_{ij}w_{ik} =&\sum\limits_{i = 1}^N(\sum\limits_{\{j, \text{s.t.} (i, j)\in G\}}w_{ij})^2\\
 \geq& \frac{1}{N}(\sum\limits_{i = 1}^N\sum\limits_{\{j, \text{s.t.} (i, j)\in G\}}w_{ij})^2 \\
 = &\frac{4 }{N}(\sum\limits_{(i, j), (k, l) \in G} w_{ij}w_{kl} ).
 \end{split}
 \end{equation*}

\begin{equation*}
\begin{split}
 \text{Var}(R_1^w-R_2^w) >0 \Leftrightarrow &\sum\limits_{\{j \in G_i\}}w_{ij} \text{ are not all equal for all } i \in [1, N],\\
 \text{Var}(q_wR_1^w+p_wR_2^w) >0 \Leftrightarrow & (N-3)S_1-S_2+\frac{2}{N-1}S_3 >0.
 \end{split}
 \end{equation*}

\section{Proof of Theorem \ref{th:decompose}} \label{Appendix:decompose}
Let $\textbf{R} = \begin{pmatrix}
  R_1^w\\ 
  R_2^w
\end{pmatrix}$, $\textbf{C} = \begin{pmatrix}
  1 & -1\\ 
  q & p
\end{pmatrix}$, $R_{\text{diff}}^w = R_1^w-R_2^w$ and $R_w^w = qR_1^w+pR_2^w$.
\begin{equation*}
\begin{split}
    S &= (\textbf{R} - E(\textbf{R} ))^T\mathbf{\Sigma}^{-1}(\textbf{R} - E(\textbf{R}))\\
    &=(\textbf{R} - E(\textbf{R} ))^T\textbf{C}^T(\textbf{C}^T)^{-1}\mathbf{\Sigma}^{-1}\textbf{C}^{-1}\textbf{C}(\textbf{R} - E(\textbf{R}))\\
    &=(\textbf{C}(\textbf{R} - E(\textbf{R} )))^T(\textbf{C}\mathbf{\Sigma}\textbf{C}^T)^{-1}(\textbf{C}(\textbf{R} - E(\textbf{R}))),\\
     \textbf{C}\mathbf{\Sigma}\textbf{C}^T & = \textbf{C}\begin{pmatrix}
  \text{Var}(R_1^w) & \text{Cov}(R_1^w, R_2^w)\\ 
  \text{Cov}(R_1^w, R_2^w) & \text{Var}(R_2^w)
\end{pmatrix}\textbf{C}^T,\\
\textbf{C}\mathbf{\Sigma}\textbf{C}^T &= \begin{pmatrix}
  \text{Var}(R_{\text{diff}}^w)& C_1\\ 
  C_1 & \text{Var}(R_w^w)
\end{pmatrix},
\end{split}
\end{equation*}
where 
\begin{equation*}
\begin{split}
\text{Var}(R_{\text{diff}}^w) &= \text{Var}(R_1^w) -2\text{Cov}(R_1^w, R_2^w) + \text{Var}(R_2^w),\\
\text{Var}(R_w^w) &= q^2\text{Var}(R_1^w) +2pq\text{Cov}(R_1^w, R_2^w) + p^2\text{Var}(R_2^w),\\
C_1 &= q\text{Var}(R_1^w) +(p-q)\text{Cov}(R_1^w, R_2^w) - p\text{Var}(R_2^w)\\
    &=D_N \left\{\frac{(N-3)(n_2-1)}{(N-2)(n_2-1)}\left(S_1+S_2 - \frac{4}{N}S_3\right)-\right.\\
    &\quad\left.\frac{(N-3)(n_1-1)}{(N-2)(n_1-1)}\left(S_1+S_2-\frac{4}{N}S_3\right)\right\} \\
    &= 0.
\end{split}
\end{equation*}

So $S_R = \frac{(R_{\text{diff}}^w - E(R_{\text{diff}}^w))^2}{\text{Var}(R_{\text{diff}}^w)} + \frac{(R_w^w - E(R_w^w))^2}{\text{Var}(R_w^w)},$
and the robust test statistic $S_R$ can be decomposed as 
$S_R = (Z^R_\text{diff})^2 + (Z^R_{w})^2$
and $\textbf{Cov}(Z^R_\text{diff},  Z^R_{w}) = 0.$

\section{Proof of Theorem \ref{th:bounds}} \label{Appendix:bounds}

For $s = 1, 2$, 
$R_j^w = \sum_{(i, j)\in G} w_{ij}I_{J_{(i, j)} = s} > \min(w_{ij})\sum_{(i, j)\in G} I_{J_{(i, j)} = s}$. 

Then $\min(w_{ij})$ is asymptotically bounded below by $1/|G|$ and $\sum_{(i, j)\in G} I_{J_{(i, j)} = s} = O(|G|)$ since $\sum_{(i, j)\in G} I_{J_{(i, j)} = s}/N$ converge to a constant related to the densities of the two samples according to Theorem 2 in \cite{Henze1999asy}.

So $\lim_{N\to\infty}\min(w_{ij})\sum_{(i, j)\in G} I_{J_{(i, j)} = s}>0$, $s = 1, 2$.

\section{Proof of Theorem \ref{th:conver}} \label{Appendix:limit_distribution}

We will use the bootstrap null distribution to prove Theorem \ref{th:conver}. Under the bootstrap null, the probability of an observation assigned to sample $\boldsymbol{X}$ is $\frac{n_X}{N}$, and the probability of an observation assigned to sample $\boldsymbol{Y}$ is $1-\frac{n_X}{N}$. When $n_x = n_1$, the bootstrap null distribution is equivalent to the permutation null. We use subscripts to denote statistics under the bootstrap null.

First, we introduce Theorem \ref{th:stein} to help prove Theorem \ref{th:conver}.

\begin{assumption}
\label{assum:a1}
[\cite{chen2005stein}, p. 17] For each $i\in J$, there exists $K_i \subset L_i \subset J$ such that $\xi_i$ is independent of $\xi_{K_i^C}$ and $\xi_{K_i}$ is independent of $\xi_{L_i^C}$.
\end{assumption}

\begin{theorem}
\label{th:stein}
[\cite{chen2005stein}, Theorem 3.4] 

Under Assumption \ref{assum:a1}, we have 
$\sup\limits_{h\in Lip(1)}|\text{E}h(W) - \text{E}h(Z)| \leq \delta,$
where $Lip(1) = \{h : R \rightarrow R\}$, $Z$ has $\mathcal{N}(0, 1)$ distribution and 
$\delta = 2\sum\limits_{i \in J}(E|\xi_i\eta_i\theta_i| + |E(\xi_i\eta_i)|E|\theta_i|)+\sum\limits_{i \in J}|E|\xi_i\eta_i^2|, $
with $\eta_i = \sum\limits_{j \in K_i}\xi_j$ and $\theta_i = \sum\limits_{j \in L_i} \xi_j$, where $K_i$ and $L_i$ are defined in Assumption \ref{assum:a1}.
\end{theorem}
Let $p_n = \frac{n_1}{N}$, $q_n = 1-\frac{n_1}{N} = \frac{n_2}{N},$
\begin{equation*}
\begin{split}
	\text{E}_B(R_1^w) &= \sum\limits_{(i, j) \in G}w_{ij}P(J_{(i, 1) = 1}) = \sum\limits_{(i, j) \in G}w_{ij}p_n^2:=\mu_1^B,\\
	\text{E}_B(R_2^w) &= \sum\limits_{(i, j) \in G}w_{ij}P(J_{(i, 1) = 2}) = \sum\limits_{(i, j) \in G}w_{ij}q_n^2:=\mu_2^B, \\
    \text{Var}_B(R_1^w) &= \sum\limits_{(i, j) \in G}w_{ij}^2p_n^2 + \sum\limits_{\substack{(i, j), (i, k) \in G\\ j\neq k}} w_{ij}w_{ik}p_n^3 + \\
    &\quad\sum\limits_{\substack{(i, j), (k, l) \in G\\ i, j, k, l \text{ all different}}} w_{ij}w_{kl}p_n^4 - (\sum\limits_{(i, j) \in G}w_{ij})^2p_n^4\\
    &=\sum\limits_{(i, j) \in G}w_{ij}^2(p_n^2 - p_n^4)+ \sum\limits_{\substack{(i, j), (i, k) \in G\\ j\neq k}} w_{ij}w_{ik}(p_n^3 - p_n^4)\\
    &=\sum\limits_{(i, j) \in G}w_{ij}^2(p_n^2 - p_n^4)+ \sum\limits_{(i, j), (i, k) \in G} w_{ij}w_{ik}(p_n^3 - p_n^4)-\\
    &\quad\sum\limits_{(i, j) \in G}w_{ij}^2(p_n^3 - p_n^4)\\
    &=\sum\limits_{(i, j) \in G}w_{ij}^2p_n^2q_n+\sum\limits_{(i, j), (i, k) \in G} w_{ij}w_{ik}p_n^3q_n\\&:=(\sigma_1^B)^2.
    \end{split}
\end{equation*}
Similarly, 
\begin{equation*}
\begin{split}
    \text{Var}_B(R_2^w) &=
    \sum\limits_{(i, j) \in G}w_{ij}^2q_n^2p_n+\sum\limits_{(i, j), (i, k) \in G} w_{ij}w_{ik}q_n^3p_n:=(\sigma_2^B)^2, \\
     \text{Cov}_B(R_1^w, R_2^w) &= \text{E}_B(R_1^wR_2^w) - \text{E}_B(R_1^w)\text{E}_B(R_2^w)\\&=
    \sum\limits_{(i, j) \in G}w_{ij}\sum\limits_{\substack{ (k, l) \in G\\ i, j, k, l \text{ all different}}} w_{kl}p_n^2q_n^2-\sum\limits_{(i, j) \in G}w_{ij}p_n^2\sum\limits_{(i, j) \in G}w_{ij}q_n^2\\&=
    -\sum\limits_{(i, j), (i, k) \in G} w_{ij}w_{ik}p_n^2q_n^2:=(\sigma_{12}^B)^2.\\
\end{split}
\end{equation*}
Let $R_\text{diff}^w = R_1^w - R_2^w$, we have
\begin{equation*}
\begin{split}
\text{E}_B(R_\text{diff}^w) & = \sum\limits_{(i, j) \in G}w_{ij}(p_n-q_n):=\mu_{diff}^B,\\
    \text{Var}_B(R_\text{diff}^w)  &=\text{Var}_B(R_1^w) +\text{Var}_B(R_2^w) -2\text{Cov}_B(R_1^w, R_2^w)\\&=p_nq_n \sum\limits_{(i, j) \in G}w_{ij}^2+\sum\limits_{(i, j), (i, k) \in G} w_{ij}w_{ik}(p_n^3q_n+q_n^3p_n+2p_n^2q_n^2)
    \\&= p_nq_n (\sum\limits_{(i, j) \in G}w_{ij}^2+\sum\limits_{(i, j), (i, k) \in G} w_{ij}w_{ik})\\
    &:=(\sigma_{diff}^B)^2.
\end{split}
\end{equation*}
Let $R_{w}^w = qR_1^w + pR_2^w$, we have
\begin{equation*}
	\begin{split}
	\text{E}_B(R_{w}^w)  &= \sum\limits_{(i, j) \in G}w_{ij}\frac{n_2^2(n_1-1)+n_1^2(n_2-1)}{N^2(N-2)}:=\mu_{w}^B, \\
    \text{Var}_B(R_{w}^w)  &=q^2\text{Var}_B(R_1^w) +p^2\text{Var}_B(R_2^w) +2pq\text{Cov}_B(R_1^w, R_2^w)\\
    &= \frac{n_1n_2(n_1-n_2)^2}{N^4(N-2)^2}\sum\limits_{(i, j), (i, k) \in G} w_{ij}w_{ik}+\\
    &\quad\frac{n_1n_2\{n_1n_2(N-4)+N\}}{N^3(N-2)^2}\sum\limits_{(i, j) \in G}w_{ij}^2\\&:=(\sigma_w^B)^2.
    \end{split}
\end{equation*}

Let, 
\begin{eqnarray*}
W_1^B &=& \frac{R_{w}^w - E_B(R_{w}^w)}{\sqrt{ \text{Var}_B(R_{w}^w)}},\\
W_2^B &=& \frac{R_\text{diff}^w - E_B(R_\text{diff}^w)}{\sqrt{ \text{Var}_B(R_\text{diff}^w)}},\\
W_3^B &=& \frac{n_X - n}{\sqrt{Np_n(1-p_n)}}.
\end{eqnarray*}

\begin{lemma}
\label{le:1}
Under conditions
\begin{enumerate}[(i)]
	\item $|G| = \mathcal{O}(N^{\alpha}), 1\leq\alpha<1.5$,
	\item $\begin{aligned}[t]
			&\sum_{(i, j) \in G}w_{ij}^2 + \sum_{(i, j), (i, k) \in G} w_{ij}w_{ik} - \frac{4}{N}\sum_{(i, j), (k, l) \in G} w_{ij}w_{kl}  \\
			=&\mathcal{O}(\sum_{(i, j) \in G}w_{ij}^2 + \sum_{(i, j), (i, k) \in G} w_{ij}w_{ik}),
	\end{aligned}$
	
    \item $\sum\limits_{(i, j)\in G}(w_{ij}|A_{(i, j)}|)^2 = o(\sum\limits_{(i, j)\in G}w_{ij}^2N^{0.5})$,

    \item $\begin{aligned}[t]
			&\sum_{(i, j)\in G}w_{ij}\sum_{(i', j')\in A_{(i, j)}}w_{i'j'}\sum_{(i'', j'')\in B_{(i, j)}}w_{i''j''}=o(\sum_{(i, j)\in G}w_{ij}^2)^{1.5},
    	\end{aligned}$
\end{enumerate}
and under the bootstrap null, $(W_1^B, W_2^B, W_3^B)$
is multivariate normal.
\end{lemma}

\begin{lemma} 
\label{le:2}
We have
\begin{itemize}
    \item $\frac{\text{Var}_B(R_{w}^w)}{\text{Var}(R_{w}^w)} \rightarrow c_1$,
    \item $\frac{\text{Var}_B(R_\text{diff}^w)}{\text{Var}(R_\text{diff}^w)} \rightarrow c_2$,
    \item $\frac{E_B(R_{w}^w) - E(R_{w}^w)}{\sqrt{\text{Var}(R_{w}^w)}}\rightarrow 0$,
    \item $\frac{E_B(R_\text{diff}^w) - E(R_\text{diff}^w)}{\sqrt{\text{Var}(R_{\text{diff}}^w)}}\rightarrow 0$,
    \item $\lim\limits_{N \to \infty}\text{Cov}(Z_w, Z_\text{diff}) = 0$,
\end{itemize}
where $c_1$ and $c_2$ are constant.
\end{lemma}

From Lemma \ref{le:1}, $(W_1^B,W_2^B|W_3^B)$ is multivariate normal under the bootstrap null. Since conditioning on $W_3^B = 0$, $(W_1^B, W_2^B|W_3^B = 0)$ and $(W_1^B, W_2^B)$ under the permutation distribution have the same distribution, and
\begin{eqnarray*}
Z^R_{w} &=& \frac{\sqrt{\text{Var}_B(R_{w}^w)}}{\sqrt{\text{Var}(R_{w}^w)}}(W_1^B + \frac{\text{E}_B(R_{w}^w) - \text{E}(R_{w}^w)}{\text{Var}_B(R_{w}^w)}),\\
Z^R_{\text{diff}} &=& \frac{\sqrt{\text{Var}_B(R_{\text{diff}}^w)}}{\sqrt{\text{Var}(R_{\text{diff}}^w)}}(W_2^B + \frac{\text{E}_B(R_{\text{diff}}^w) - \text{E}(R_{\text{diff}}^w)}{\text{Var}_B(R_{\text{diff}}^w)}),
\end{eqnarray*}
with Lemma \ref{le:2}, we conclude that $Z^R_{w}$ and $Z^R_{\text{diff}}$ are Gaussian under the permutation distribution.

\subsection{Proof of Lemma \ref{le:1}}

We first show $(W_1^B, W_2^B, W_3^B)$ is multivariate Gaussian under the bootstrap null distribution, which is equivalent to showing that $W=a_1W_1^B+a_2 W_2^B+a_3 W_3^B$ is asymptotically Gaussian distributed for each $(a_1, a_2, a_3) \in \mathbb{R}^3$ such that $\text{Var}_B(W)>0$ by Cramer-Wold theorem.

Let the index set $J = \{(i, j)\in G\} \bigcup \{1, 2, \dots, N\}$,
\begin{equation*}
	\begin{split}
    \xi_{(i, j)} =& a_1\left(\frac{w_{ij}\frac{m-1}{N-2}I(J_{(i, j) = 1})+w_{ij}\frac{n-1}{N-2}I(J_{(i, j) = 2})}{\sigma_w^B} - \frac{ w_{ij}\frac{n^2(m-1)+m^2(n-1)}{N^2(N-2)}}{\sigma_w^B}\right)+\\
    &a_2\frac{w_{ij}I(J_{(i, j) = 1})-w_{ij}I(J_{(i, j) = 2}) -(w_{ij}(p_n-q_n))}{\sigma_{diff}^B}, \\
    \xi_i =& a_3\frac{I(g_i = 0) - p_n}{\sqrt{Np_n(1-p_n)}}.
    \end{split}
\end{equation*}

Let, $a = max(|a_1|, |a_2|, |a_3|)$, $\sigma = \text{min}(\sigma_w^B, \sigma_{diff}^B)$, $\sigma_0 = \sqrt{Np_n(1-p_n)}$. $\sigma^2$ is at least of order $\sum\limits_{(i, j)\in G}w_{ij}^2$, $\sigma_0 = O(N^{0.5})$. Then $|\xi_{(i, j)}| \leq \frac{2a}{w_{ij}\sigma}$, $|\xi_i| \leq \frac{a}{\sigma_0}$ and $W = \sum_{j \in J}\xi_j$. 

For $(i, j) \in J$, let 
\begin{equation*}
	\begin{split}
&A_{(i, j)} = \{(i, j)\} \cup \begin{aligned}[t]\{&(i', j')\in G: (i', j') \text{ and } (i, j) \text{ share a node}\},\end{aligned}\\
&B_{(i, j)} = A_{(i, j)} \cup\begin{aligned}[t]
\{&(i'', j'')\in G: \exists(i', j')\in A_{(i, j)},\\
&\text{ s.t. } (i', j')\text{ and } (i'', j'') \text{ share a node}\},
\end{aligned}\\
&K_{(i, j)} = A_{(i, j)} \cup \{i, j\},\\
&L_{(i, j)} = B_{(i, j)} \cup \{\text{nodes in }A_{(i, j)}\}.
 \end{split}
\end{equation*}
For $i \in \{1, 2, ..., N\},$ let
\begin{align*}
&G_i = \{(i, j) \in G\},\\
&G_{i, 2} = \{(i, j) \in G\}\cup \begin{aligned}[t]\{&(i'', j'')\in G: \exists(i', j')\in G_i,\\
&\text{ s.t. }(i', j')\text{ and } (i'', j'') \text{ share a node}\},\end{aligned} \\
&K_i = G_i \cup \{i\},\\
&L_j = G_{i, 2} \cup \{\text{nodes in }G_i\}.
\end{align*}
For $j \in J$, let $\eta_j = \sum_{k \in K_j} \xi_k$ and $\theta_j = \sum_{k \in L_j} \xi_k$ .
$$sup_{h \in Lip(1)}|E_Bh(W) - Eh(Z)|\leq \delta \text{ for } Z\sim N(0, 1),$$
where $\delta =\frac{1}{\sqrt{\text{Var}_B(W)}}\big(2\sum_{j\in J}(E_B|\xi_j\eta_j\theta_j|+E_B(\xi_j\eta_j)E_B|\theta_j|) + \sum_{j\in J}E_B|\xi_j\eta_j^2|\big)$, according to Theorem \ref{th:stein}. 
For $j \in \{1, 2, ..., N\}$,
\begin{equation*}
	\begin{split}
	\eta_j&=\sum_{k \in K_j} \xi_k = \xi_i + \sum_{(i', j')\in G_i}\xi_{(i', j')}\leq \frac{a}{\sigma_0}+\frac{2a}{\sigma}\sum_{(i', j')\in G_i}w_{i'j'},\\
	\theta_j&=\sum_{k \in L_j} \xi_k =\sum_{\text{nodes in }G_i} \xi_i + \sum_{(i', j')\in G_{i, 2}}\xi_{(i', j')}\leq 2\frac{a|G_i|}{\sigma_0}+\frac{2a}{\sigma}\sum_{(i', j')\in G_{i, 2}}w_{i'j'}.
	\end{split}
\end{equation*}
So,
\begin{equation*}
	\begin{split}
	 &\quad2\sum_{j \in \{1, 2, ..., N\}}(E_B|\xi_j\eta_j\theta_j|+E_B(\xi_j\eta_j)E_B|\theta_j|) + \sum_{j \in \{1, 2, ..., N\}}E_B|\xi_j\eta_j^2|\\&\leq5\frac{a^3}{\sigma_0}(\frac{1}{\sigma_0}+\frac{2}{\sigma}\sum_{(i', j')\in G_i}w_{i'j'})(2\frac{|G_i|}{\sigma_0}+\frac{2}{\sigma}\sum_{(i', j')\in G_{i, 2}}w_{i'j'}).
\end{split}
\end{equation*}

For $(i, j) \in G$,
\begin{equation*}
	\begin{split}
	\eta_{(i, j)}&=\sum_{k \in K_{(i, j)}} \xi_k = \xi_i+\xi_j + \sum_{(i', j')\in A_{(i, j)}}\xi_{(i', j')}\\
	&\leq \frac{2a}{\sigma_0}+\frac{2a}{\sigma}\sum_{(i', j')\in A_{(i, j)}}w_{i'j'}, \\
	\theta_{(i, j)}&=\sum_{k \in L_{(i, j)}} \xi_k =\sum_{\text{nodes in }A_{(i, j)}} \xi_i + \sum_{(i', j')\in B_{(i, j)}}\xi_{(i', j')}\\&\leq 2\frac{a|A_{(i, j)}|}{\sigma_0}+\frac{2a}{\sigma}\sum_{(i', j')\in B_{(i, j)}}w_{i'j'}.
	\end{split}
\end{equation*}
So,
\begin{equation*}
	\begin{split}
    &\quad 2\sum_{(i, j) \in G}\left(E_B|\xi_{(i, j)}\eta_{(i, j)}\theta_{(i, j)}|+E_B(\xi_{(i, j)}\eta_{(i, j)})E_B|\theta_{(i, j)}|\right)\\
    & \quad + \sum_{(i, j) \in G}E_B|\xi_{(i, j)}\eta_{(i, j)}^2|\\&
    \leq
    5\frac{2aw_{ij}}{\sigma }(\frac{2a}{\sigma_0}+\frac{2a}{\sigma}\sum_{(i', j')\in A_{(i, j)}}w_{i'j'})(2\frac{a|A_{(i, j)}|}{\sigma_0}+\frac{2a}{\sigma}\sum_{(i', j')\in B_{(i, j)}}w_{i'j'})\\&=
   40\frac{a^3w_{ij}}{\sigma }(\frac{1}{\sigma_0}+\frac{1}{\sigma}\sum_{(i', j')\in A_{(i, j)}}w_{i'j'})(\frac{|A_{(i, j)}|}{\sigma_0}+\frac{1}{\sigma}\sum_{(i', j')\in B_{(i, j)}}w_{i'j'}).
\end{split}
\end{equation*}

Then we have
\begin{equation*}
	\begin{split}
    \delta 
    \leq &[\sum\limits_{(i, j)\in G} 40\frac{a^3w_{ij}}{\sigma }(\frac{1}{\sigma_0}+\frac{1}{\sigma}\sum_{(i', j')\in A_{(i, j)}}w_{i'j'})(\frac{|A_{(i, j)}|}{\sigma_0}+\frac{1}{\sigma}\sum_{(i', j')\in B_{(i, j)}}w_{i'j'})+\\
    &\sum\limits_{i=1}^N 5\frac{a^3}{\sigma_0}(\frac{1}{\sigma_0}+\frac{2}{\sigma}\sum_{(i', j')\in G_i}w_{i'j'})(2\frac{|G_i|}{\sigma_0}+\frac{2}{\sigma}\sum_{(i', j')\in G_{i, 2}}w_{i'j'})]\frac{1}{\sqrt{\text{Var}_B(W)}}.
\end{split}
\end{equation*}

If we want $\delta \rightarrow 0$ as $N\rightarrow \infty$, we need the following conditions to hold:
\begin{enumerate}
    \item[(1)] $\begin{aligned}[t]
    &\sum\limits_{i=1}^N\sum\limits_{(i', j')\in G_i}w_{i'j'}\sum\limits_{(i'', j'')\in G_{i, 2}}w_{i''j''} = o(\sum\limits_{(i, j)\in G}w_{ij}^2N^{0.5}),
    \end{aligned}$
    \item[(2)] $\sum\limits_{i=1}^N\sum\limits_{(i', j')\in G_i}w_{i'j'}|G_i| = o((\sum\limits_{(i, j)\in G}w_{ij}^2)^{0.5}N)$,
    \item[(3)] $\sum\limits_{i=1}^N\sum\limits_{(i', j')\in G_{i,2}}w_{i'j'}= o((\sum\limits_{(i, j)\in G}w_{ij}^2)^{0.5}N)$,
    \item[(4)] $\sum\limits_{i=1}^N|G_i| = o(N^{1.5})$,
    \item[(5)] $\sum\limits_{(i, j)\in G}w_{ij}|A_{(i, j)}| = o((\sum\limits_{(i, j)\in G}w_{ij}^2)^{0.5}N)$,
    \item[(6)] $\sum\limits_{(i, j)\in G}w_{ij}\sum\limits_{(i', j')\in B_{(i, j)}}w_{i'j'}= o(\sum\limits_{(i, j)\in G}w_{ij}^2N^{0.5})$,
    \item[(7)] $\begin{aligned}[t]
    &\sum\limits_{(i, j)\in G}w_{ij}|A_{(i, j)}|\sum\limits_{(i', j')\in A_{(i, j)}}w_{i'j'}= o(\sum\limits_{(i, j)\in G}w_{ij}^2N^{0.5}),
    \end{aligned}$
    \item[(8)]  $\begin{aligned}[t]
			&\sum\limits_{(i, j)\in G}w_{ij}\sum\limits_{(i', j')\in A_{(i, j)}}w_{i'j'}\sum\limits_{(i'', j'')\in B_{(i, j)}}w_{i''j''}=o(\sum\limits_{(i, j)\in G}w_{ij}^2)^{1.5}.
    	\end{aligned}$
\end{enumerate}

We need conditions:
\begin{enumerate}[(i)]
    \item $|G| = \mathcal{O}(N^{\alpha}), 1\leq\alpha<1.5$,
    \item $\sum\limits_{(i, j)\in G}(w_{ij}|A_{(i, j)}|)^2 = o(\sum\limits_{(i, j)\in G}w_{ij}^2N^{0.5})$,
    \item $\sum\limits_{(i, j)\in G}w_{ij} = o((\sum\limits_{(i, j)\in G}w_{ij}^2)^{0.5}N)$,
    \item $\begin{aligned}[t]
			&\sum\limits_{(i, j)\in G}w_{ij}\sum\limits_{(i', j')\in A_{(i, j)}}w_{i'j'}\sum\limits_{(i'', j'')\in B_{(i, j)}}w_{i''j''}=o(\sum\limits_{(i, j)\in G}w_{ij}^2)^{1.5},
    	\end{aligned}$
    \item $\begin{aligned}[t]
    &\sum\limits_{(i, j) \in G}w_{ij}^2 + \sum\limits_{(i, j), (i, k) \in G} w_{ij}w_{ik} - \frac{4}{N}\sum\limits_{(i, j), (k, l) \in G} w_{ij}w_{kl} \\
    =& \mathcal{O}(\sum\limits_{(i, j) \in G}w_{ij}^2 + \sum\limits_{(i, j), (i, k) \in G} w_{ij}w_{ik}).
    \end{aligned}$
\end{enumerate} 

Since 
\begin{equation*}
	\begin{split}
    \sum\limits_{(i', j')\in A_{(i, j)}}w_{i'j'}\leq d_i \frac{1}{d_i} + d_j \frac{1}{d_j} + \frac{1}{\max(d_i, d_j)} = O(1) = \mathcal{O}(|A_{(i, j)}|w_{ij}).
\end{split}
\end{equation*}
$\sum_{(i', j')\in A_{(i, j)}}w_{i'j'} = \mathcal{O}(|A_{(i, j)}|w_{ij})$, and condition (7) holds according to (ii).

Let $\gamma_{G_i}$ denotes the vertex set of $G_i/\{i\}$,
\begin{equation*}
	\begin{split}
    \sum\limits_{i=1}^N\sum\limits_{(i', j')\in G_i}w_{i'j'}\sum\limits_{(i'', j'')\in G_{i, 2}}w_{i''j''}&\leq  \sum\limits_{i=1}^N\sum\limits_{(i', j')\in G_i}w_{i'j'}\sum\limits_{j \in \gamma_{G_i}}\sum\limits_{(i'', j'')\in G_j}w_{i''j''}\\&=
    \sum\limits_{i=1}^N\sum\limits_{j \in \gamma_{G_i}}\sum\limits_{(i', j')\in G_i}w_{i'j'}\sum\limits_{(i'', j'')\in G_j}w_{i''j''}\\&=2 \sum\limits_{(i, j)\in G}\sum\limits_{(i', j')\in G_i}w_{i'j'}\sum\limits_{(i'', j'')\in G_j}w_{i''j''}\\&\leq 2 \sum\limits_{(i, j)\in G}(\sum\limits_{(i', j')\in A_{(i, j)}}w_{i'j'})^2 \\&= \mathcal{O}(\sum\limits_{(i, j)\in G}w_{ij}|A_{(i, j)}|\sum\limits_{(i', j')\in A_{(i, j)}}w_{i'j'}).
\end{split}
\end{equation*}
So condition (7) implies condition (1).

By Cauchy-Schwarz inequality and (ii)
\begin{equation*}
	\begin{split}
    \sum\limits_{(i, j)\in G}w_{ij}|A_{(i, j)}| &\leq \sqrt{\sum\limits_{(i, j)\in G}w_{ij}^2|A_{(i, j)}|^2|G| }\\&=o((\sum\limits_{(i, j)\in G}w_{ij}^2)^{0.5}N^{0.25})|G|^{0.5}.
\end{split}
\end{equation*}
So (i) ensures that condition (5) holds. 
$$ \sum\limits_{(i, j)\in G}\sum\limits_{(i', j')\in A_{(i, j)}}w_{i'j'}= \mathcal{O}(\sum\limits_{(i, j)\in G}w_{ij}|A_{(i, j)}| )$$
\begin{equation*}
	\begin{split}
  \sum\limits_{(i, j)\in G}\sum\limits_{(i', j')\in A_{(i, j)}}w_{i'j'} &=\sum\limits_{(i, j)\in G}(\sum\limits_{(i', j')\in G_i}w_{i'j'}+\sum\limits_{(i'', j'')\in G_j}w_{i''j''} - w_{ij})\\&=\sum\limits_{i = 1}^N\sum\limits_{j \in \gamma_{G_i}}\sum\limits_{(i', j')\in G_j}w_{i'j'}-\sum\limits_{(i, j)\in G}w_{ij}\\&=\sum\limits_{i=1}^N\sum\limits_{(i', j')\in G_i}w_{i'j'}|G_i|-\sum\limits_{(i, j)\in G}w_{ij}.
\end{split}
\end{equation*}
According to conditions (5) and (iii), condition (2) holds.
\begin{equation*}
G_{i, 2} \subset \bigcup\limits_{j\in\gamma_{G_i}}G_j,
\end{equation*}
\begin{equation*}
	\begin{split}
   \sum\limits_{i=1}^N\sum\limits_{(i', j')\in G_{i,2}}w_{i'j'} &\leq \sum\limits_{i=1}^N\sum\limits_{j\in\gamma_{G_i}}\sum\limits_{(i', j')\in G_j}w_{i'j'}\\& = \sum\limits_{(i, j)\in G}(\sum\limits_{(i', j')\in G_i}w_{i'j'}+\sum\limits_{(i'', j'')\in G_j}w_{i''j''})\\&\leq2\sum\limits_{(i, j)\in G}\sum\limits_{(i', j')\in A_{(i, j)}}w_{i'j'} = \mathcal{O}(\sum\limits_{(i, j)\in G}w_{ij}|A_{(i, j)}|).
\end{split}
\end{equation*}
So condition (5) implies condition (3).
\begin{equation*}
	\begin{split}
    \sum\limits_{(i, j)\in G}w_{ij}\sum\limits_{(i', j')\in B_{(i, j)}}w_{i'j'} &\leq \sum\limits_{(i, j)\in G}w_{ij}\sum\limits_{(i', j')\in A_{(i, j)}}\sum\limits_{(i'', j'')\in A_{(i', j')}}w_{i''j''}\\&=\sum\limits_{(i, j)\in G}w_{ij}(\sum\limits_{(i', j')\in A_{(i, j)}}w_{i'j'})^2 \\&= \mathcal{O}(\sum\limits_{(i, j)\in G}w_{ij}|A_{(i, j)}|\sum\limits_{(i', j')\in A_{(i, j)}}w_{i'j'}).
\end{split}
\end{equation*}
So condition (7) implies condition (6).

Finally, since $(\sum\limits_{(i, j)\in G}w_{ij})^2 \leq |G|\sum\limits_{(i, j)\in G}w_{ij}^2$ , 
\begin{equation*}
	\begin{split}
	\sum\limits_{(i, j)\in G}w_{ij} &=o( |G|^{0.5}(\sum\limits_{(i, j)\in G}w_{ij}^2)^{0.5})=  o((\sum\limits_{(i, j)\in G}w_{ij}^2)^{0.5}N),
	\end{split}
\end{equation*}
if condition (i) is satisfied. 

So we need conditions (i), (iii), (iv), (v).

\subsection{Proof of Lemma \ref{le:2}}

\begin{equation*}
	\begin{split}
	&\text{Var}_B(R_{w}^w)\\
	 = &\frac{n_1n_2(n_1-n_2)^2}{N^4(N-2)^2}\sum\limits_{(i, j), (i, k) \in G} w_{ij}w_{ik}+\frac{n_1n_2\{n_1n_2(N-4)+N\}}{N^3(N-2)^2}\sum\limits_{(i, j) \in G}w_{ij}^2\\
	 = &\mathcal{O}(\sum\limits_{(i, j) \in G}w_{ij}^2 ),
	\end{split}
\end{equation*} 
since \begin{equation*}
	\begin{split}
	\sum\limits_{(i, j), (i, k) \in G} w_{ij}w_{ik}&\leq(\sum\limits_{(i, j) \in G}w_{ij})^2 = o(\sum\limits_{(i, j) \in G}w_{ij}^2N^2).\end{split}
\end{equation*}
\begin{equation*}
	\begin{split}
	\text{Var}(R_{w}^w) &=  \frac{n_1n_2(n_1-1)(n_2-1)}{N(N-1)(N-2)(N-3)}\big\{\sum\limits_{(i, j) \in G}w_{ij}^2 - \\
	&\quad\frac{1}{N-2}(\sum\limits_{(i, j) \in G}w_{ij}^2 +\sum\limits_{(i, j), (i, k) \in G} w_{ij}w_{ik} - \frac{4}{N}\sum\limits_{(i, j), (k, l) \in G} w_{ij}w_{kl}) -\\
	 &\quad\frac{2}{N(N-1)}\sum\limits_{(i, j), (k, l) \in G} w_{ij}w_{kl}\big\} \\
	&=  \mathcal{O}(\sum\limits_{(i, j) \in G}w_{ij}^2 ).
	\end{split}
\end{equation*}

So, $\lim_{N \to \infty}\frac{\text{Var}_B(R_{w}^w)}{\text{Var}(R_{w}^w)}= c_1$, where $c_1$ is a constant.

\begin{equation*}
	\begin{split}
	\lim\limits_{N \to \infty}\frac{\text{Var}_B(R_\text{diff}^w)}{\text{Var}(R_\text{diff}^w)} = &\lim\limits_{N \to \infty}\left(\sum\limits_{(i, j) \in G}w_{ij}^2 + \sum\limits_{(i, j), (i, k) \in G} w_{ij}w_{ik} \right)/\\
	&\left(\sum\limits_{(i, j) \in G}w_{ij}^2 + \sum\limits_{(i, j), (i, k) \in G} w_{ij}w_{ik} - \frac{4}{N}\sum\limits_{(i, j), (k, l) \in G} w_{ij}w_{kl}\right)  \\
	&= c_2,
	\end{split}
\end{equation*}
where $c_2$ is a constant, according to condition (v).

Since $E_B(R_{w}^w) - E(R_{w}^w) = \frac{n_1n_2}{N^2(N-1)}\sum_{(i, j)\in G}w_{ij}$,
$$\lim\limits_{N \to \infty}\frac{E_B(R_{w}^w) - E(R_{w}^w)}{\sqrt{\text{Var}(R_{w}^w)}} =\lim\limits_{N \to \infty}\frac{1}{N}\frac{\sum\limits_{(i, j)\in G}w_{ij}}{c_3\sqrt{\sum\limits_{(i, j) \in G}w_{ij}^2}},$$
where $c_3$ is a constant. 

From condition (iii) $\sum_{(i, j)\in G}w_{ij} = o((\sum_{(i, j)\in G}w_{ij}^2)^{0.5}N)$, 
$$\lim_{N \to \infty}\frac{E_B(R_{w}^w) - E(R_{w}^w)}{\sqrt{\text{Var}(R_{w}^w)}} =0.$$

Since $E_B(R_\text{diff}^w) - E(R_\text{diff}^w) = 0$, 
$$\lim_{N \to \infty}\frac{E_B(R_\text{diff}^w) - E(R_\text{diff}^w)}{\sqrt{\text{Var}(R_\text{diff}^w)}} =0.$$

We still need to show $\lim_{N \to \infty}\text{Cov}(Z_w, Z_\text{diff}) = 0$.
\begin{equation*}
	\begin{split}
	\text{Cov}(Z_w, Z_\text{diff}) &= \frac{E(R_{w}^wR_\text{diff}^w) -E(R_{w}^w)E(R_\text{diff}^w) }{\sqrt{\text{Var}(R_{w}^w)\text{Var}(R_\text{diff}^w)}}, \\
    E(R_{w}^wR_\text{diff}^w) &= S_3[q\frac{n_1^2(n_1-1)^2}{N^2(N-1)^2}- p\frac{n_2^2(n_2-1)^2}{N^2(N-1)^2}+\\
    &\quad (p-q)\frac{n_1n_2(n_1-1)(n_2-1)}{N^2(N-1)^2}]\\
    &=\frac{(n_1-1)(n_2-1)(n_1-n_2)}{N(N-1)(N-2)}S_3,\\
    E(R_{w}^w)E(R_\text{diff}^w) &= S_3[(\frac{n_1-n_2}{N})(\frac{n_1n_2-N+1}{(N-1)(N-2)})],
\end{split}
\end{equation*}
where $S_3 = \sum_{(i, j), (k, l) \in G} w_{ij}w_{kl}$.
\begin{equation*}
	\begin{split}
	\lim_{N \to \infty}E(R_{w}^wR_\text{diff}^w) &= \sum_{(i, j), (k, l) \in G} w_{ij}w_{kl}p_nq_n(p_n-q_n),\\
	\lim_{N \to \infty}E(R_{w}^w)E(R_\text{diff}^w) &= \sum_{(i, j), (k, l) \in G} w_{ij}w_{kl}p_nq_n(p_n-q_n).
	\end{split}
\end{equation*}

So $\lim_{N \to \infty}(E(R_{w}^wR_\text{diff}^w) -E(R_{w}^w)E(R_\text{diff}^w)) = 0$.

\bibliographystyle{abbrvnat}
\bibliography{bibliography}

\end{document}